\documentclass[fleqn,usenatbib]{mnras}

\usepackage{newtxtext,newtxmath}

\usepackage[T1]{fontenc}
\usepackage{ae,aecompl}
\usepackage{url}

\usepackage{graphicx}	
\usepackage{amsmath}	
\usepackage{amssymb}	
\usepackage{float}

\newcommand{\nup}{$\nu_{\rm peak}^S$}

\newcommand{\swift}{{\it Swift}}

\usepackage{lineno}

\title[Dissecting the region around IceCube-170922A]{Dissecting the region around IceCube-170922A: 
the blazar TXS\,0506+056 as the first cosmic neutrino source}

\author[P. Padovani et al.]{P. Padovani$^{1}$, P. Giommi$^{2,3,4}$, E. Resconi$^5$,
T. Glauch$^5$, B. Arsioli$^{6,7}$, N. Sahakyan$^8$, \newauthor M. Huber$^5$\\
$^{1}$European Southern Observatory, Karl-Schwarzschild-Str. 
2, D-85748 Garching bei M\"unchen, Germany\\
$^{2}$Agenzia Spaziale Italiana, ASI, via del Politecnico s.n.c., I-00133 Roma Italy \\
$^{3}$Institute for Advanced Studies, Technische 
Universit{\"a}t M{\"u}nchen, Lichtenbergstrasse 2a, 
D-85748 Garching bei M\"unchen, Germany\\
$^{4}$ICRANet, Piazzale della Repubblica,10 - 65122, Pescara, 
Italy\\
$^{5}$Technische Universit{\"a}t M{\"u}nchen, Physik-Department, 
James-Frank-Str. 1, D-85748 Garching bei M{\"u}nchen, Germany\\
$^{6}$Instituto de F\'{i}sica Gleb Wataghin, UNICAMP, R. S\'{e}rgio 
Buarque de Holanda 777, 
13083-859 Campinas, Brazil \\
$^{7}$ICRANet-Rio, CBPF, Rua Dr. Xavier Sigaud 150, 22290-180 URCA, 
Rio de Janeiro, Brazil \\
$^{8}$ICRANet-Armenia, Marshall Baghramian Avenue 24a, 0019 
Yerevan, Republic of Armenia\\
}

\date{Accepted 2018 July 9. Received 2018 July 6; in original form 2018 May 17}

\pubyear{2018}

\begin{document}
\label{firstpage}
\pagerange{\pageref{firstpage}--\pageref{lastpage}}
\maketitle

\begin{abstract}
We present the dissection in space, time, and energy of the region around
the IceCube-170922A neutrino alert.  This study is motivated by: (1) the
first association between a neutrino alert and a blazar in a flaring state,
TXS\,0506+056; (2) the evidence of a neutrino flaring activity during
2014\,--\,2015 from the same direction; (3) the lack of an accompanying
simultaneous $\gamma$-ray enhancement from the same counterpart; (4) the contrasting flaring activity
of a neighbouring bright $\gamma$-ray source, the blazar PKS\,0502+049, during 2014\,--\,2015. Our study makes use of
multi-wavelength archival data accessed through Open Universe tools and
includes a new analysis of {\it Fermi}-LAT data.  We find that
PKS\,0502+049 contaminates the $\gamma$-ray emission region at low energies
but TXS\,0506+056 dominates the sky above a few GeV. TXS\,0506+056, which
is a very strong (top percent) radio and $\gamma$-ray source, is in a high
$\gamma$-ray state during the neutrino alert but in a low though hard
$\gamma$-ray state in coincidence with the neutrino flare. Both states can
be reconciled with the energy associated with the neutrino emission and, in
particular during the low/hard state, there is evidence that TXS\,0506+056
has undergone a hadronic flare with very important implications for blazar
modelling.  All multi-messenger diagnostics reported here support a single
coherent picture in which TXS\,0506+056, a very high energy $\gamma$-ray 
blazar, is the only counterpart
of all the neutrino emissions in the region and therefore the most
plausible first non-stellar neutrino and, hence, cosmic ray source.
\end{abstract}



\section{Introduction}\label{sec:Introduction}

The IceCube Neutrino Observatory at the South Pole\footnote{\url{http://icecube.wisc.edu}} has recently reported the detection of a number of high-energy astrophysical 
neutrinos\footnote{In this paper neutrino means both neutrino and anti-neutrino.} 
\citep{2013PhRvL.111b1103A,ICECube13,ICECube14,ICECube15_1}. These include 82 high-energy starting 
events collected over six years \citep{ICECube17_2}, which are inconsistent with a purely 
atmospheric origin with 
a significance greater than 6.5\,$\sigma$.
The IceCube signal, including the 34 through-going charged current $\nu_\mu$ from the northern sky \citep{Aartsen2015,ICECube15_2,Aartsen2016,ICECube17_1}, is still compatible with an isotropic distribution. 
The origin of the IceCube neutrinos is presently unknown \citep[see, e.g.][and references therein, for 
a comprehensive discussion]{Ahlers_2015} although various hints consistently point to blazars as one of the most probable candidates, as described below. 

The Large Area Telescope (LAT) on-board the {\it Fermi Gamma-ray Space Telescope} is a pair-conversion telescope sensitive to high energy  photons with energies from 20\,MeV to greater than 300\,GeV \citep{atwood}
that has been surveying the $\gamma$-ray sky for the past almost 10 years. The constant monitoring and archiving of all-sky $\gamma$-ray data permits unprecedented investigations of variable sources.

To explore the complexity of the multi-wavelength sky, we make use of innovative tools that are under development 
within ``Open Universe'', a new initiative under the auspices of the United Nations Committee On the 
Peaceful Uses of Outer Space (COPUOS).
The goal of Open Universe is to stimulate a large increase of the accessibility and usability of space 
science data in all sectors of society from the professional scientific community, to universities, 
schools, museums, and citizens. 
A web portal of the Open Universe initiative, developed at the Italian Space Agency, 
is available at \url{openuniverse.asi.it}.
In this paper we make use of software, 
such as the VOU-BLAZAR tool, which has been specifically designed to 
identify blazars based on multi-frequency 
information in large error regions, and spectral energy distribution (SED) animations. 

Blazars are active galactic nuclei \citep[AGN; see][for a recent AGN
review]{Padovani_2017} having a relativistic jet that is seen at a small angle with 
respect to the line of 
sight. The jet contains charged particles moving in a magnetic field emitting 
non-thermal radiation over 
the entire electromagnetic spectrum 
\citep{UP95,Padovani_2017}. 
Since the energy distribution of these particles can significantly differ from object 
to object, the 
electromagnetic emission exhibits a wide range of intensity levels and spectral slopes 
across the spectrum. This results in observational properties that depend strongly on 
the energy band where blazars are discovered. 
In a series of papers \cite{giommibsv1,giommibsv2,padovanibsv1,giommibsv3} 
proposed a new blazar paradigm \citep[but see][for an alternative scenario]
{Ghisellini_2017}
based on dilution by the jet and the host galaxy, minimal assumptions on the physical 
properties of the 
non-thermal jet emission, and unified schemes. By means of detailed Monte Carlo 
simulations, it was shown that this scenario is 
consistent with the complex observational properties of blazars as we know them in all 
parts of the electromagnetic spectrum.

The possibility that blazars could be the sources of high-energy neutrinos has been investigated by many authors, even long before the IceCube detections
\citep[e.g.][]{mannheim95,halzen97,mueckeetal03,Pad_2014,Petro_2015,tav15}.

\cite{Padovani_2016} have correlated the second catalogue of hard {\it Fermi}-LAT sources (2FHL, $E>50$\,GeV, 
\citealt{2FHL}) and other catalogues, with the publicly available high-energy neutrino sample detected by IceCube. The chance probability 
of association of 2FHL high-energy peaked blazars (HBL/HSP, i.e. sources with the peak of the synchrotron emission \nup~$> 10^{15}$~Hz\footnote{Blazars can be further divided into low (LBL:  
\nup~$<10^{14}$~Hz) and intermediate (IBL: $10^{14}$~Hz$<$ \nup~$<
10^{15}$~Hz) energy peaked sources.}; \citealt{padgio95}) with IceCube events
was 0.4 per cent, which becomes 1.4 per cent ($2.2\,\sigma$) by 
evaluating the impact of trials \citep{Resconi_2017}. This 
hint appears to be strongly dependent on $\gamma$-ray flux. The corresponding fraction 
of the IceCube signal explained by HBL is however only $\sim 10 - 20$ per cent,
which agrees with the results of \cite{Aartsen2017,ICECube17_3}, who by searching for 
cumulative neutrino emission from blazars in the second {\it Fermi}-LAT AGN 
\citep[2LAC;][]{Fermi2LAC} 
and other catalogues (including also the 2FHL), have constrained the 
maximum contribution of known
blazars to the observed astrophysical neutrino flux to $< 27$ per cent.

High-energy astrophysical neutrinos originate in cosmic ray interactions providing a natural link with high-energy and possibly ultrahigh-energy cosmic ray (UHECR) detection. 
\cite{Resconi_2017} have presented a hint of a connection between HBL, IceCube neutrinos, and UHECRs ($E \ge 52 \times 10^{18}$ eV) with a
probability $\sim 0.18$ per cent ($2.9\,\sigma)$ after compensation for all
the considered trials. Even in this case, HBL can account only for $\approx 10$ per cent of the UHECR signal.

None of the possible neutrino counterparts in \cite{Padovani_2016} and \cite{Resconi_2017} are tracks, as they are 
all cascade-like events\footnote{The topology of IceCube detections can be 
broadly classified in two types: (1) cascade-like, characterized by a compact spherical 
energy deposition, which can only be reconstructed with a spatial resolution $\approx 15^{\circ}$; 
(2) track-like, defined by a dominant linear topology from the induced muon, 
with positions known typically within one degree or less.}. This indicates that
by using tracks we are still limited in  sensitivity to the HBL neutrino signal. 
Although tracks trace only about 1/6 of the astrophysical signal for a 
flavour ratio $\nu_{\rm e}:\nu_\mu:\nu_\tau = 1:1:1$, 
standard neutrino cross-sections, and IceCube event selection efficiencies,
after a long enough exposure a track IceCube signal from blazars should also 
start to appear. This is of great interest, because false (random) associations of 
tracks with a blazar are unlikely due to the better defined position of this 
event-class with respect to cascades.

Recently \cite{Lucarelli_2017} have found  a transient $\gamma$-ray ($> 100$\,MeV) 
{\it AGILE} source positionally coincident with an 
IceCube track with a post-trial significance $\sim 4\,\sigma$ and possibly 
associated with an HBL. 
However, no other space missions nor ground observatories have reported any 
detection of transient 
emission consistent with this event.

The most probable hint of an association ($3 - 3.5\,\sigma$) reported so far 
\citep{icfermi} between an IceCube 
astrophysical neutrino and an extragalactic object 
is that of the neutrino IceCube-170922A and the radio bright ($\sim$ 1\,Jy at 5\,GHz) and $\gamma$-ray flaring BL Lac 
object TXS\,0506+056 (also known as 5BZB\,J0509+0541, 2FHL J0509.5+0541, and 3FGL\,J0509.4+0541).
Moreover, IceCube has reported in \cite{iconly}  
an independently observed 3.5\,$\sigma$ excess of neutrinos from the direction of TXS\,0506+056 between October 2014 and February 2015 providing further indication of a high-energy neutrino association. 

The presence of several non-thermal objects including variable blazars 
within 80 arc-minutes of IceCube-170922A (that is within the size of the {\it 
Fermi}-LAT point spread function [PSF\footnote{The Fermi PSF for this analysis event selection has been determined using the \texttt{gtpsf} Fermi Science Support Center tool.}], {which is $\sim 2.8^{\circ}$ [95 per cent containment] at $E = 1$\,GeV)
makes the $\gamma$-ray emission from this area quite complex, with possible source confusion.  
Different objects could in fact contribute to the overall $\gamma$-ray flux at different levels in a 
time-dependent manner. For this reason, we report here on what we have called the {\it dissection} of 
the 
region around IceCube-170922A  taking into account the fact that all sources present in the area could 
be in principle contributors to the neutrino emission observed in 2014--2015 and in 2017.
We use innovative software tools that exploit all the publicly available multi-frequency data to  study in 
detail the area around the position of IceCube-170922A at all energies and in the time domain, 
together 
with a very careful  analysis of  the $\gamma$-ray emission, providing a wide perspective in space 
and time. 
 
Section 2 describes the multi-messenger data we used, while Section 3 puts them all together 
to study the relevant sources in the area, their $\gamma$-ray light curves and SEDs. 
Section 4 gives our results, which are discussed in Section 5. Section 6 summarizes our conclusions. 
We use a $\Lambda$CDM cosmology with Hubble constant $H_0 = 70$ km
s$^{-1}$ Mpc$^{-1}$, matter density $\Omega_{\rm m,0} = 0.3$, and dark energy density $\Omega_{\Lambda,0} = 0.7$. 

\section{Multi-messenger data analysis}\label{sec:MMData}

\subsection{Neutrino data}
\subsubsection{The IceCube-170922A alert event}
The high-energy upward-going muon IceCube-170922A reported by IceCube through 
a Gamma-ray Coordinates Network   
Circular on MJD~58018 \citep[September 22, 2017;][]{2017GCN.21916....1K} originates from a 
neutrino with 
$E_\nu \sim 290$\,TeV, which is probably of astrophysical origin. 
The best-fit reconstructed position is right 
ascension (RA) $77.43^{+0.95}_{-0.65}$ and declination (Dec) 
$+5.72^{+0.50}_{-0.30}$ (deg, J2000, 90 per cent containment region:
\citealt{icfermi}). 
No other high-energy neutrino passing the same selection of alert-like events 
has been observed from this direction from 2010 onwards until today. 

\cite{icfermi} have derived the coincidence
probability as a measure of the likelihood that a neutrino alert like
IceCube-170922A is correlated by chance with a flaring blazar, considering
the large number of known $\gamma$-ray sources and the modest number of
neutrino alerts. This has been done through several hypothesis tests
covering a range of assumptions on the spatial and temporal signal
distribution and neutrino emission scenarios. The derived probability was
trial-corrected by multiplying the p-value by 51, which correspond to the number of alerts
issued by IceCube (10) plus the 41 inspected archival events, which would
have triggered alerts if the realtime system had been operational. The final post-trial
coincidence probability ranges between $2.5 \times 10^{-4}$ and $1.3 \times 10^{-3}$ ($3 - 3.5\,\sigma$).

\subsubsection{The IceCube neutrino flare in 2014--2015}\label{sec:nu_flare}

In contrast with the neutrino alert discussed above, which is a single event identified 
in real time and satisfying stringent selection criteria, we define here a neutrino flare as 
a statistically significant ($> 3\,\sigma$) accumulation of neutrinos
coming from a specific direction over a well-defined time period.

As reported in \cite{iconly}, the investigation of the historical 
9.5 years of IceCube data at the position of TXS\,0506+056 revealed 
an excess with a post-trial coincidence probability $2 \times 10^{-4}$ ($3.5\,\sigma$)
over 110$^{+35}_{-24}$ days corresponding to a $\nu_\mu$ 
fluence at 100\,TeV of $2.1^{+0.9}_{-0.7} \times 10^{-4}$\,TeV  
cm$^{-2}$, spectral index $\gamma\,=\,2.1\pm0.2$, 
and an energy range (68 per cent) between 32\,TeV and 3.6\,PeV ($3.6 \times 10^{15}$\,eV)\footnote{
We use here the results obtained for the Gaussian time window
because they are the most significant.}. This could be interpreted as the first evidence of high-energy neutrino emission from the 
direction of a 
known source and comes from an excess of $\sim 13$ neutrino events between 
MJD\,56949 and 57059 (October 19, 2014 -- February 6, 2015).

\subsection{Radio and optical monitoring data}

The radio (15\,GHz) and optical (V$_{\rm mag}$) data have been taken from the Owens Valley Radio Observatory (OVRO\footnote{\url{http://www.astro.caltech.edu/ovroblazars/}}) database, the Catalina Real time Transient Survey 
(CRTS\footnote{\url{http://crts.caltech.edu}}) and from the All Sky Automatic Survey (ASAS\footnote{\url{http://www.astrouw.edu.pl/asas/}}; \citealt{asas}) online services.

\subsection{X-ray and optical/UV data}\label{sec:X-rayData}

\begin{figure*}
\includegraphics[height=8.5cm]{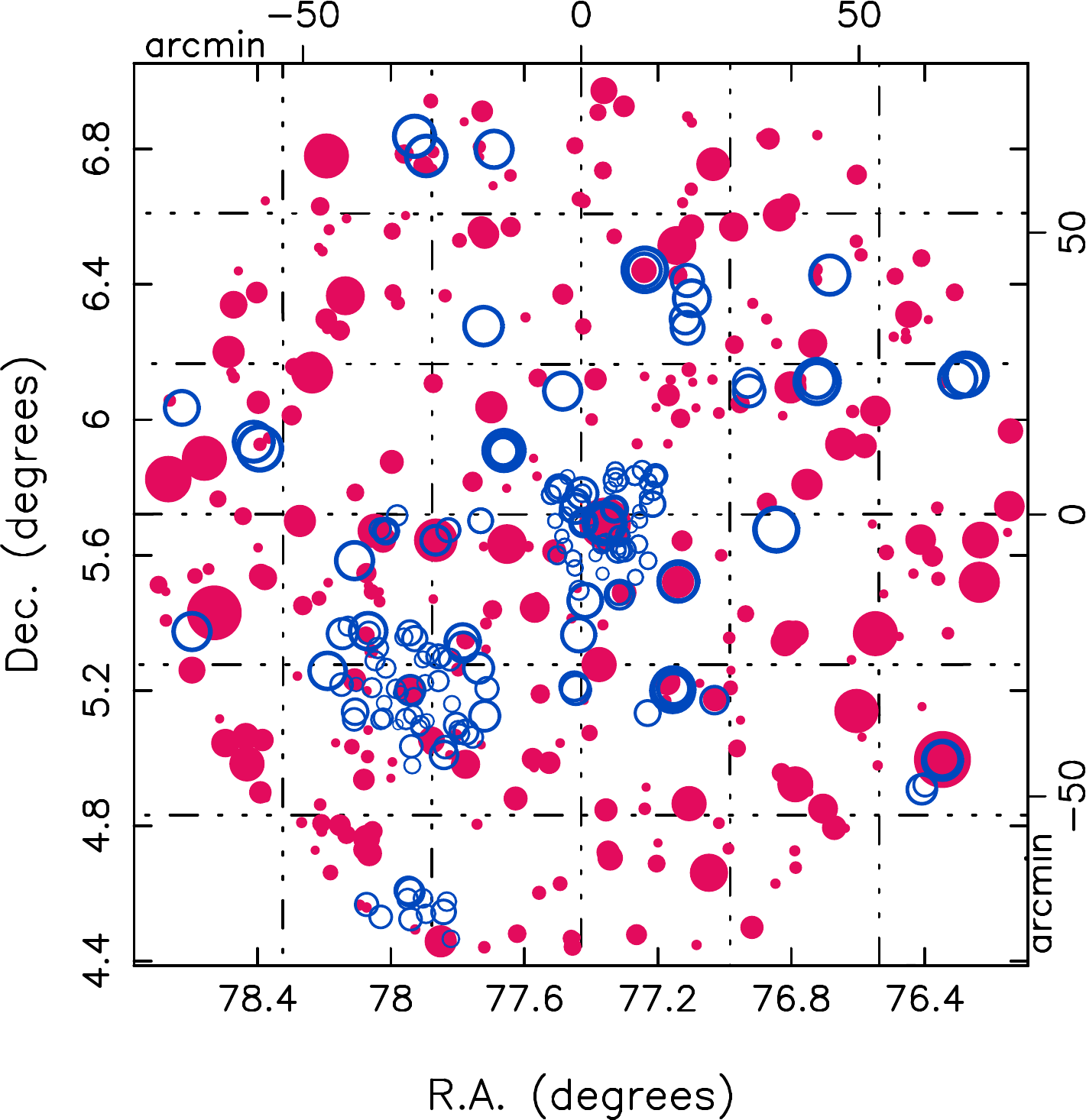}
\includegraphics[height=8.8cm]{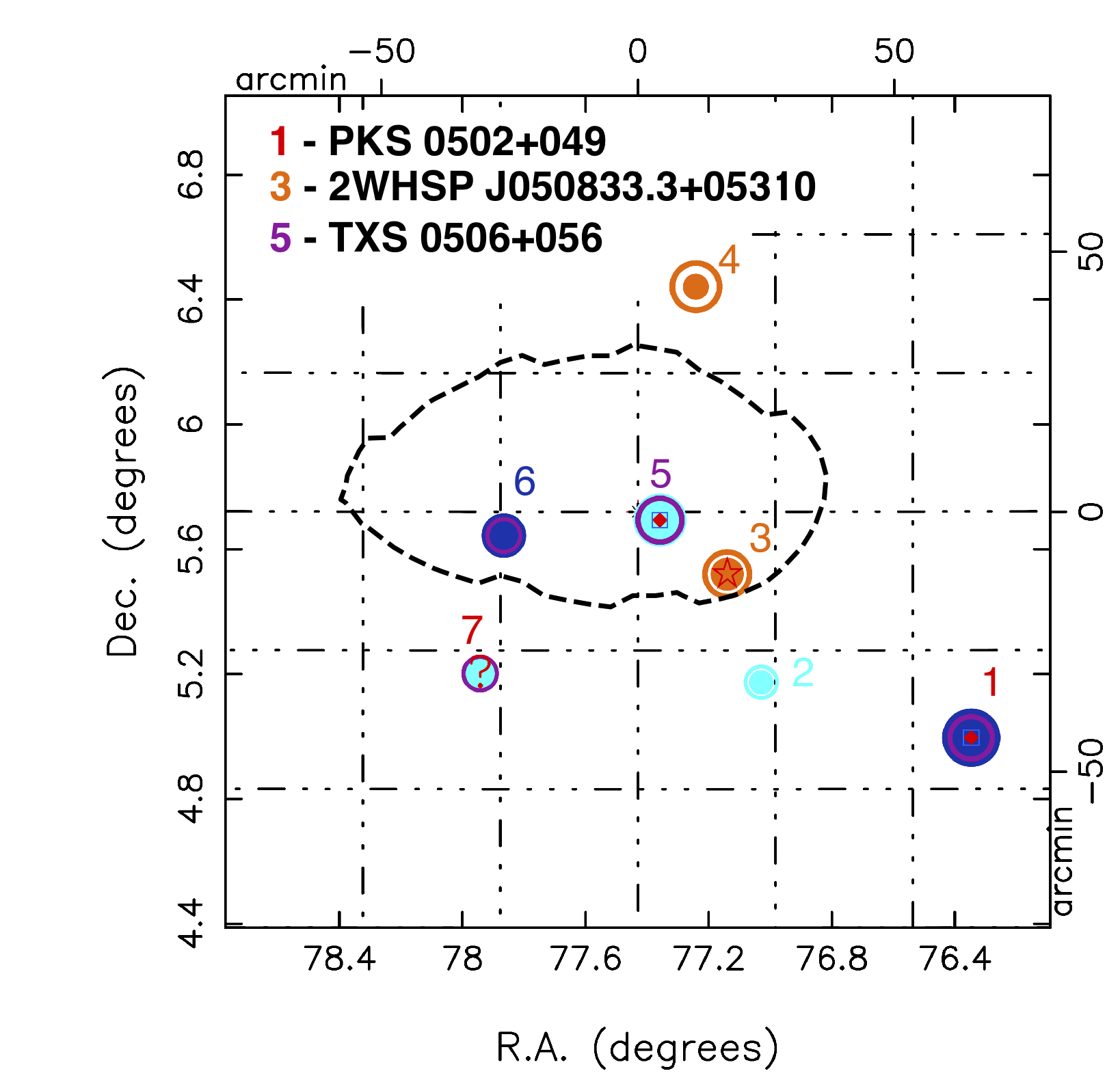}
\caption{{\bf Left}: Radio and X-ray sources within 80 arc-minutes of   
the position of IceCube-170922A. Symbol diameters are proportional 
to source intensity. Radio sources appear as red filled 
circles, X-ray sources as open blue circles. 
{\bf Right}: Known and candidate blazars around IceCube-170922A as detected by a tool developed within the framework of the Open Universe initiative as described in the text.
Dark blue circles represent LBL type candidates, that is sources with flux ratio 
in the range observed in the sample of LBL blazars of the latest edition of the BZCAT catalogue  
\citep{5bzcat}, cyan symbols are for IBL type candidates, and orange symbols are for HBL candidates. 
Known blazars are also marked by a red diamond if they are included 
in the BZCAT catalogue or a star if they are part of the 2WHSP 
sample (see text for sources n. 2, 4, 6, and 7). 
The diameters of filled and open circles are proportional to radio 
flux density and X-ray flux, respectively. 
The dashed line shows the 90 per cent error contour of IceCube-170922A. The 
localization of the {\it Fermi} sources is such that 
the error ellipses are smaller than the size of the symbols.}
\label{fig:map1}
\end{figure*}

The {\it Neil Gehrels Swift} observatory \citep{swift} carried out a total of 92 observations within 80 arc-minutes of IceCube-170922A. Of these, 35 were performed before the arrival of the neutrino and were mostly pointed at the nearby 
flat spectrum radio quasar (FSRQ) PKS\,0502+049. The remaining pointings have been carried out either as a Target of Opportunity follow up mapping the error region of IceCube-170922A a few hours after the event \citep{keivani}, or as part of 
the monitoring program of the blazar TXS\,0506+056 triggered by the IceCube neutrino alert.

We have analyzed all 92 X-Ray Telescope \citep[XRT;][]{burrows}  observations using the latest version of the \swift~data reduction software (HEADAS 6.22) applying standard procedures.  This led to the detection of 251 X-ray sources, which 
were combined with those of existing X-ray catalogues to build Fig.\ref{fig:map1} (left). 
X-ray spectral data were used together with the available multi-frequency data to assemble the SEDs of all interesting sources 
in the field (see Sect. \ref{sec:relevant}).

All optical and UV data of the \swift~Ultra-Violet and Optical telescope \citep[UVOT;][]{Roming} for TXS\,0506+056 and PKS\,0502+049 were analyzed using the SSDC online interactive archive\footnote{\url{http://www.asdc.asi.it}}. 

The {\it NuSTAR} hard X-ray observatory \citep{nustar} was pointed twice, on September 29 and October 19 2017, at TXS\,0506+056 following the detection of IceCube-170922A. A few days after the observations the data were made openly available. We have analyzed these data sets using the online analysis tool of the SSDC archives following the standard procedure. In both observations the spectral shape shows a sharp hardening at about $4 - 5$~keV.  

\subsection{$\gamma$-ray data}\label{sec:Gamma-rayData}

We used publicly available {\it Fermi}-LAT Pass 8 (with the P8R2\_SOURCE\_V6 instrument 
response functions) data acquired in the period from  
August 4, 2008 to February 10, 2018 and followed the standard 
procedures suggested by the 
{\it Fermi}-LAT team. Only 
the events with a high probability of being photons (evclass = 128, 
evtype = 3 [FRONT+BACK]) in the energy range of 100\,MeV -- 300\,GeV from a region of interest (ROI) 
defined as a circle of radius $12^\circ$ centred at the $\gamma$-ray 
position of TXS\,0506+056 (RA, Dec = 77.364, 5.699) were analyzed. 
We removed a possible contamination from the Earth limb by cutting 
out all the events with zenith angle $> 90^\circ$ and only used the 
time intervals in which the data acquisition of the spacecraft was 
stable (DATA\_QUAL>0 \&\& LAT\_CONFIG==1). 
Consistently with the event selection we used the standard Galactic (\texttt{gll\_iem\_v06}\footnote{\url{https://fermi.gsfc.nasa.gov/ssc/data/access/lat/BackgroundModels.html}}) and  
isotropic (\texttt{iso\_P8R2\_SOURCE\_V6\_v06}$^{11}$) models to describe the diffuse background emissions.


\subsubsection{Test Statistic maps}
To evaluate the presence of relevant $\gamma$-ray signatures around the arrival direction of 
IceCube-170922A, we built test statistics (TS) maps of the region \citep{Mattox}. 
The test statistic for all the $\gamma$-ray analysis in this paper is defined as 
\begin{equation}
TS = 2\times\left[\rm{ln} \mathcal{L}\,(\rm{source}) - \rm{ln} \mathcal{L}\,(\rm{no source})\right],
\label{eq:ts}
\end{equation}
where $\mathcal{L}\,(\rm{source})$ represents the nested likelihood of the data given a specific 
source 
hypothesis and $\mathcal{L}\,(\rm{no source})$ the likelihood of the background 
model. In our TS maps the signal hypothesis of a $\gamma$-ray 
point source is tested against a background model consisting of a diffuse 
Galactic and a diffuse isotropic 
component, as well as all the the {\it Fermi}-LAT third source 
catalogue \citep[3FGL;][]{3FGL} sources that lie outside the region 
of the TS map. While the point source is modeled using a power-law with free 
normalization and spectral index, 
the parameters of the background sources remain fixed.
According to Wilks' theorem the test-statistic distributions 
follows a $\mathcal{X}^2$ distribution with two degrees of freedom. 
Hence TS values of 30 and 8 are equivalent to a 5 and $2\,\sigma$ significance, respectively.
We centred our 80$\times$80 
arc-minute maps at IceCube-170922A  and used an equally spaced grid with 0.05$^\circ$ step size in 
right 
ascension and declination. For each of the grid points the {\it Fermi} Science Tools unbinned 
likelihood 
analysis tool \texttt{gtlike} was used to maximize the likelihoods in eq. \eqref{eq:ts} with respect to 
the free parameters.
We built TS maps for different time windows and energy cuts, resolving the $\gamma$-ray activity during 
the periods of the neutrino detections and for energy thresholds 1\,GeV, 2\,GeV, and 5\,GeV. Our choices
are driven by the the need for sufficiently high space resolution to distinguish different $\gamma$-ray sources (as the PSF decreases with energy) and by 
the fact that, given the possible neutrino production scenarios, 
we are mostly interested in the photons at the highest energies. 

\begin{figure}
\includegraphics[width = 0.95\columnwidth]{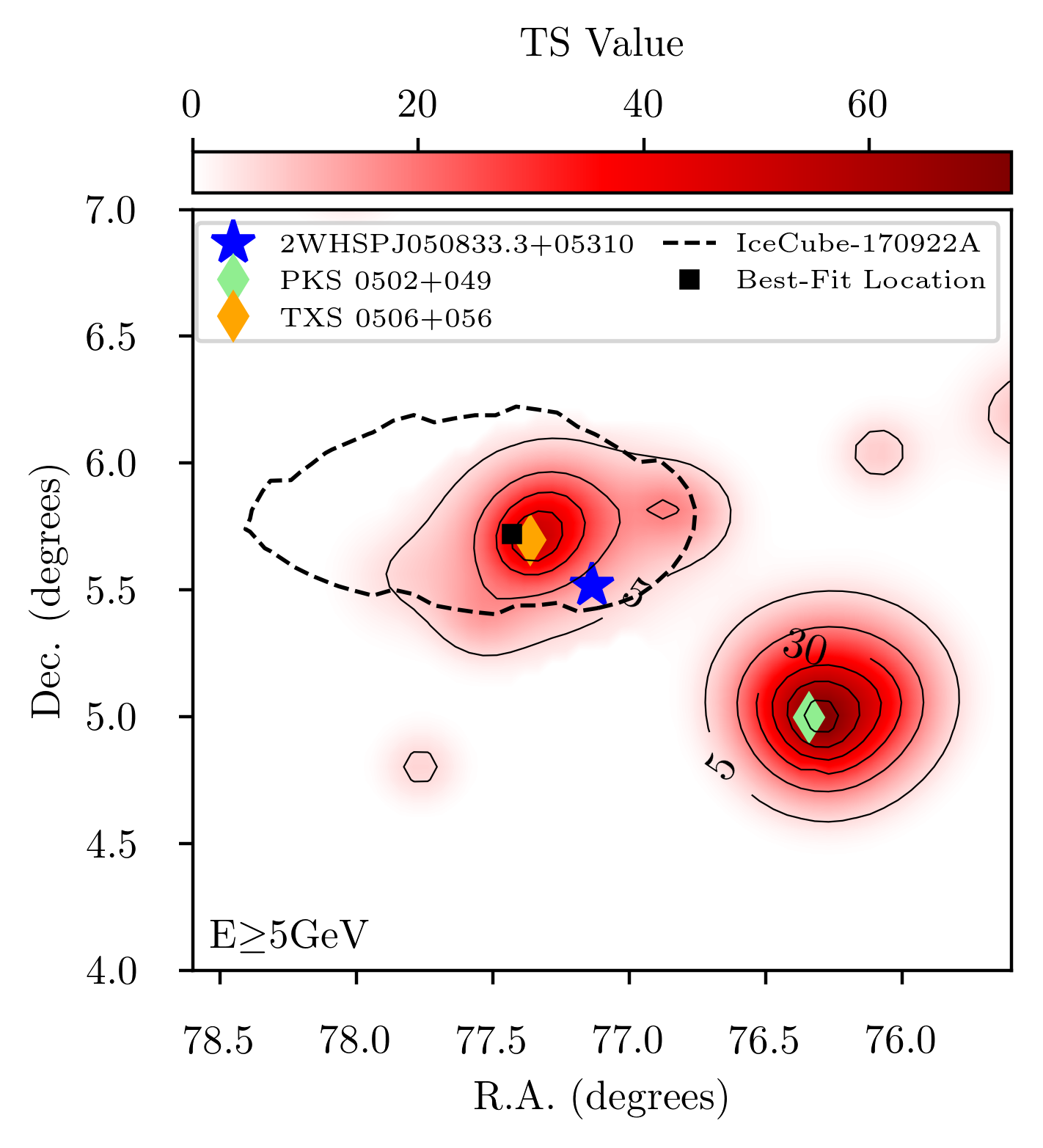}
\caption{{\it Fermi} TS map based on photons with energy larger than 
5\,GeV between MJD 55762 and 55842 (July 20 -- October 8, 2011). The dashed line shows 
the 90 per cent error contour of IceCube-170922A with best-fit location indicated as a 
black square. 
The solid contour lines connect points with the same TS value. Linear interpolation 
has been applied between the bins.}
\label{fig:fermi55800-5gev}
\end{figure}

\begin{figure*}
\includegraphics[height=230pt]{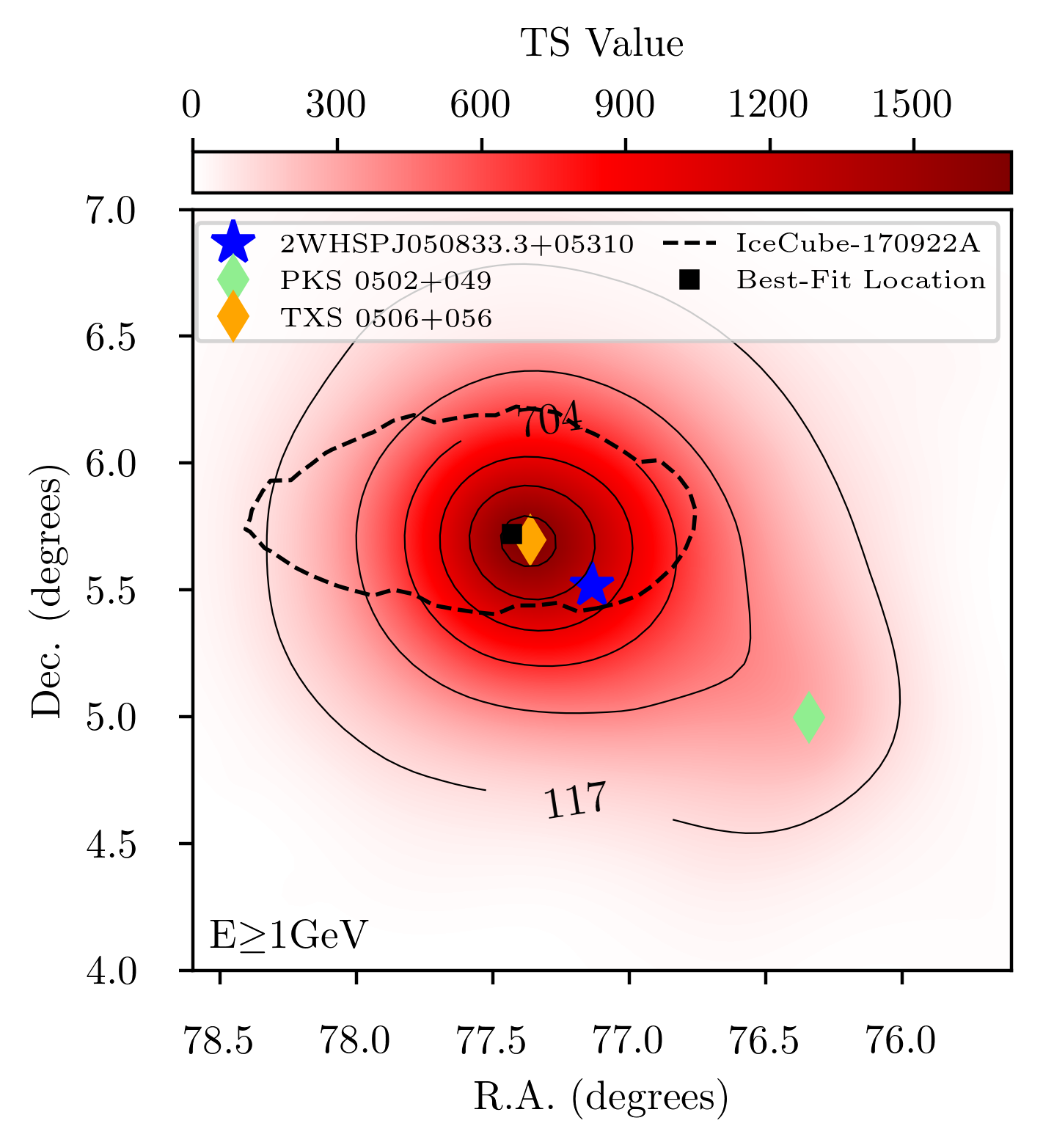}
\includegraphics[height=230pt]{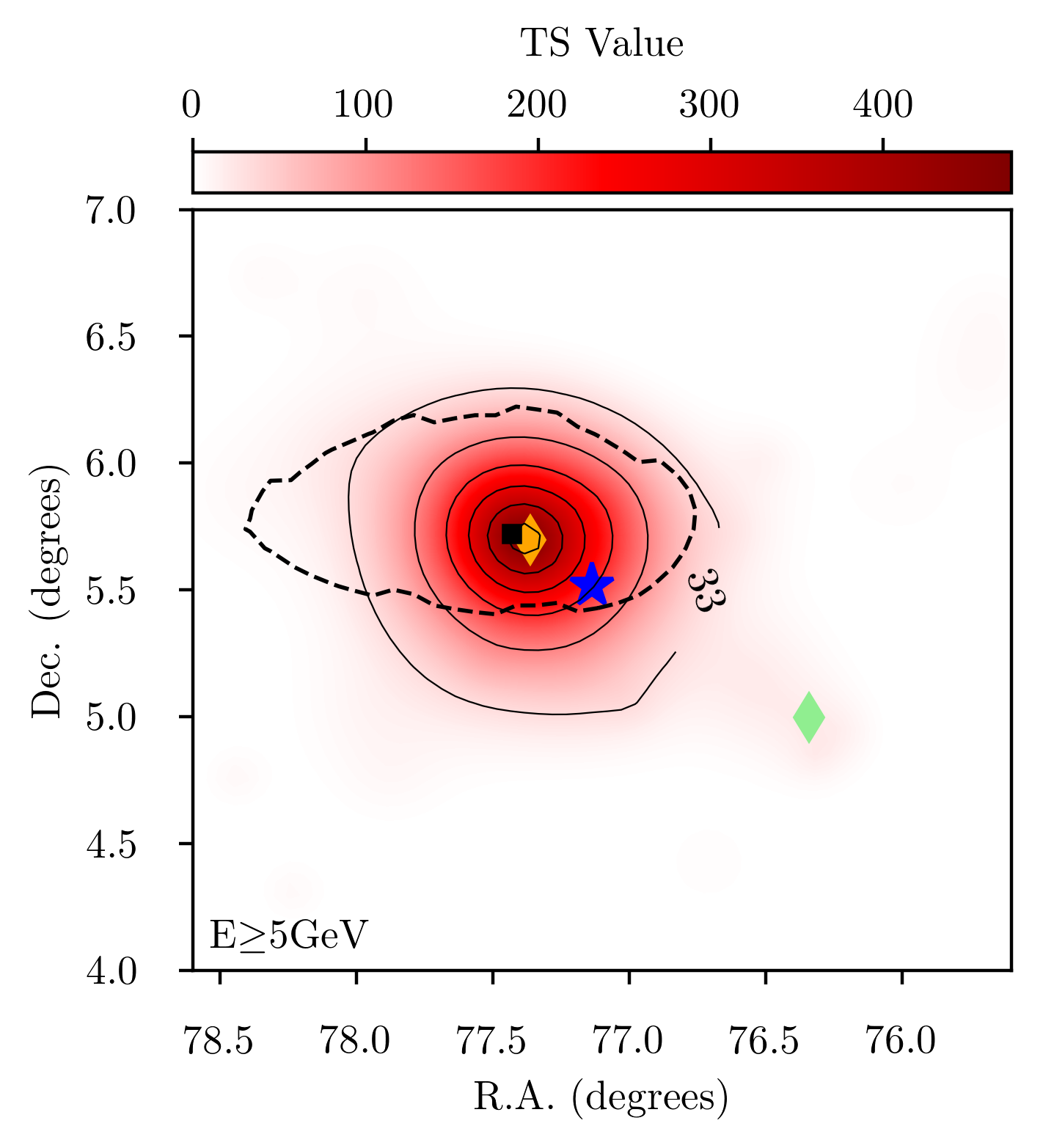}
\caption{{\it Fermi} TS map based on photons with energy larger than 
1\,GeV (left) and 5\,GeV (right) between MJD\,57908 and 58018 
(June 4 -- September 22, 2017). In this period TXS\,0506+056 
is in outburst and dominates the field. See Fig. \ref{fig:fermi55800-5gev} for more information.}
\label{fig:fermi57900-1-5gev}
\end{figure*}

\begin{figure*}
\includegraphics[height=195pt]{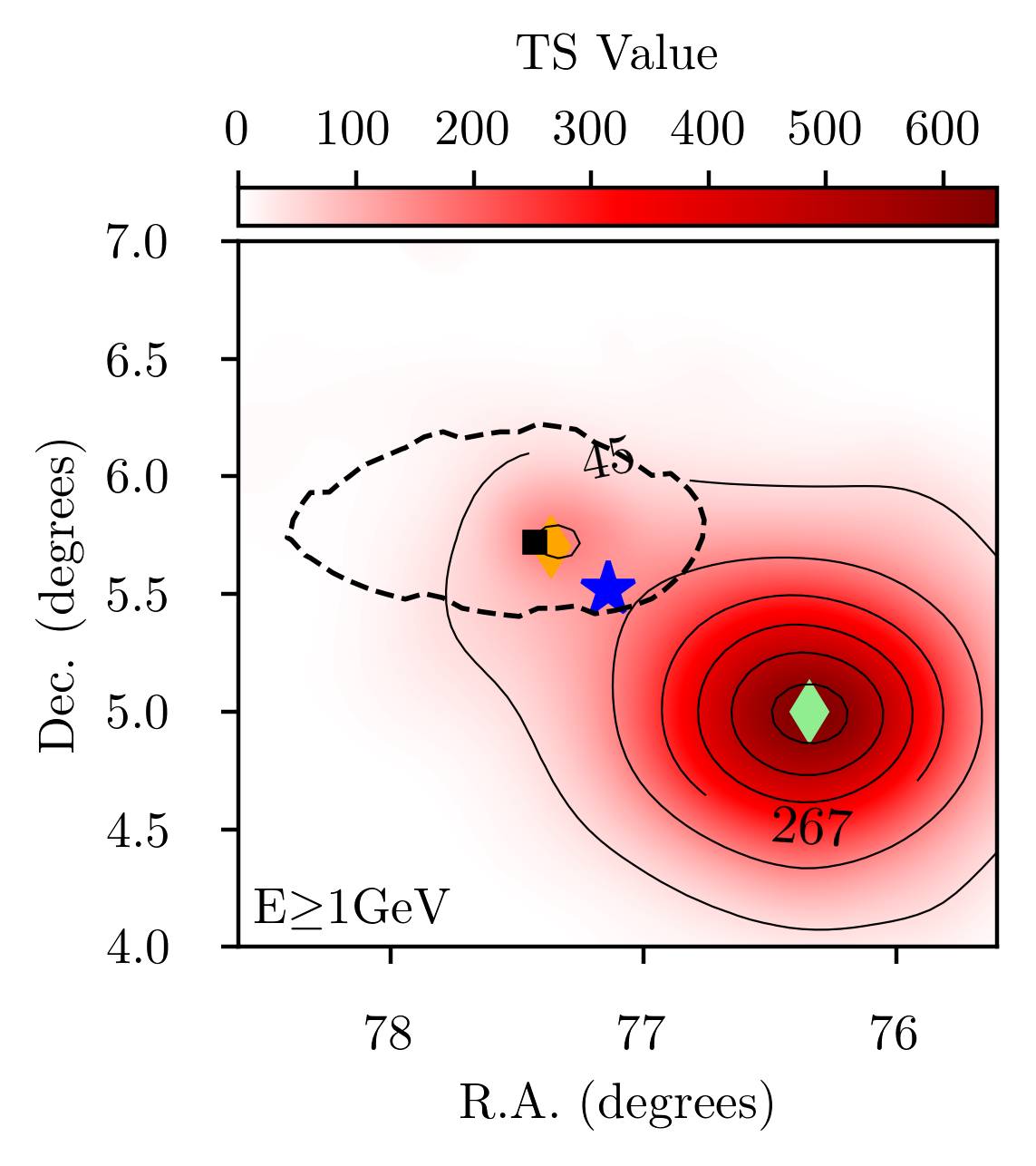}
\includegraphics[height=195pt]{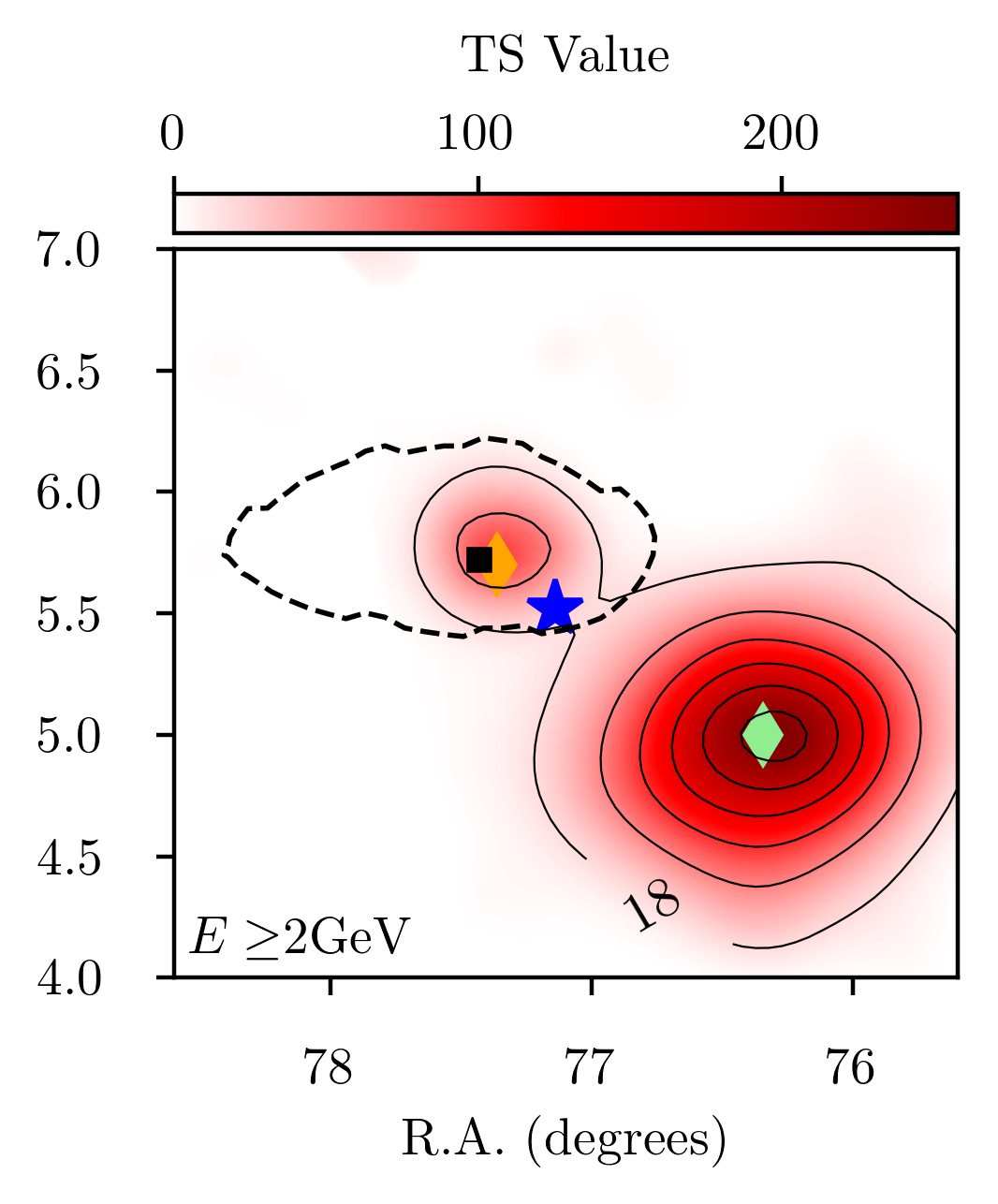}
\includegraphics[height=195pt]{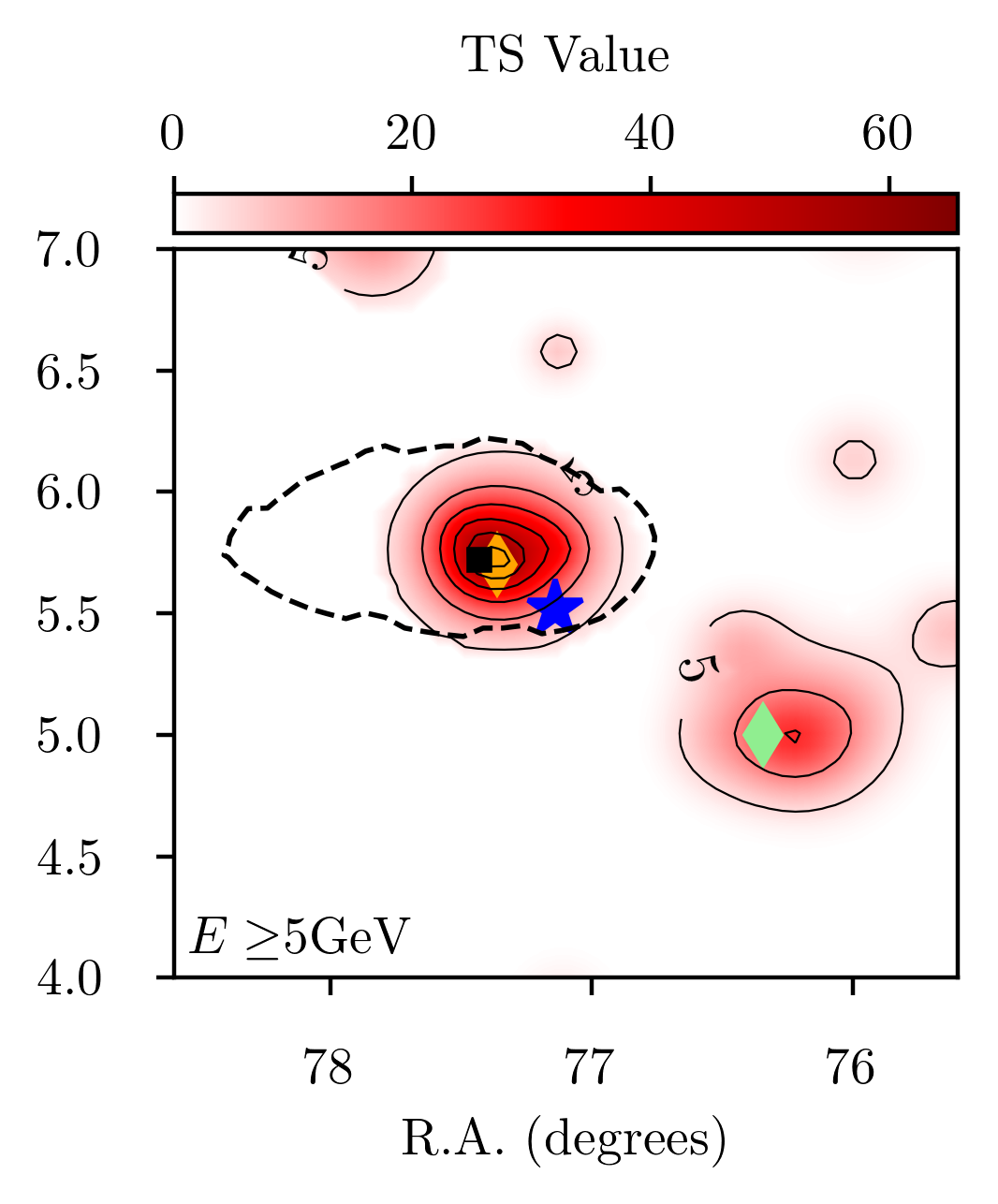}
\caption{{\it Fermi} TS map based on photons with energy larger than 1, 2 and 5\,GeV
between MJD\,56949 and 57059
(October 19, 2014 -- February 6, 2015). 
In this period the brightest source was PKS\,0502+049 at
lower energies while TXS\,0506+056 dominated at higher energies. See Fig. \ref{fig:fermi55800-5gev} for more information.}
\label{fig:fermi57100-1-5gev}
\end{figure*}

\subsubsection{Light curve and photon index}\label{sec:lc_pi}
$\gamma$-ray flux and photon index variations were investigated using light curves generated with a 
fixed time binning of 55 (28\footnote{This difference is due to 
the fact that the photon statistics for PKS\,0502+049 is better than 
for TXS\,0506+056 for the chosen energy thresholds. The bin 
sizes used have been chosen in order to have enough photon statistics 
in the majority of the bins but also to avoid any fine tuning.}) days for TXS\,0506+056 (PKS\,0502+049) and with an adaptive 
binning method with a constant relative flux uncertainty \citep{Lott_2012}. 
In the fixed time binned light curve, photons in the 2 -- 100\,GeV and 
0.1 -- 100\,GeV ranges were used for TXS\,0506+056 and PKS\,0502+049, respectively (see below). In the adaptive 
binning light curve analysis for PKS\,0502+049 the fluxes are 
computed above an optimum energy of $E_{\rm min} = 214$\,MeV in order to reach the required constant relative flux 
uncertainty of 15 per cent. 
For all the light curves the flux normalization and photon index of the target source are determined by 
applying the \texttt{gtlike} tool in each time bin. The model 
file, describing the ROI, contains point sources from the 3FGL 
catalogue within ROI+$5^\circ$ from the target, as 
well as the Galactic and isotropic $\gamma$-ray background models. It is generated using the user 
contributed \texttt{make3FGLxml.py\footnote{\url{https://fermi.gsfc.nasa.gov/ssc/data/analysis/user/}}} tool. 
For the case of TXS\,0506+056 the chosen energy threshold of 2\,GeV efficiently removes any source confusion
(see Fig. \ref{fig:fermi57100-1-5gev} and section \ref{sec:LC}), hence only this source is left free for the fit. 
The resulting light curve is robust against mis-modelling and strong time-variability of the other sources in the ROI. The light curve of 
PKS\,0502+049, on the other hand, is derived by reaching lower energies,
since the majority of the photons are below 1\,GeV \citep{3FGL}.
In the source model both sources are fitted at the same time.
The diffuse background 
components are fixed to their nine year value, since they are not expected to vary on the time scales of this analysis.  
Additionally, since we are using short integration times, we model 
PKS\,0502+049 with a power-law.

\section{Putting it all together}

\subsection{Relevant astronomical sources in the region of IceCube-170922A}\label{sec:relevant}

We have searched for non-thermal emission in blazar-like sources in the vicinity of 
IceCube-170922A using a tool that is being developed in the 
framework of the United Nations 
Open Universe initiative\footnote{\url{http://openuniverse.asi.it}} 
using Virtual Observatory protocols. This was used to find 
sources in all available lists of radio and X-ray emitters. 
Fig.~\ref{fig:map1} (left) shows a plot 
of the 637 radio and X-ray sources present 
in those catalogues or which were detected in dedicated 
\swift~observations within 80 arc-minutes of the position of IceCube-170922A. 
Of these, only 7 emit in both bands and are all blazar-like in their X-ray-to-radio 
flux ratio, as determined from the BZCAT  \citep{5bzcat} and 2WHSP  \citep{2whsp} 
samples. These are shown in Fig.\,\ref{fig:map1} (right). 

Three known objects, i.e. TXS\,0506+056 (an IBL/HBL\footnote{A fit to the SED around the time
of the neutrino alert gives \nup~$\sim 10^{15}$~Hz, making this an IBL/HBL source.} at $z = 0.3365$: \citealt{Paiano_2018}), 
PKS\,0502+049 (also known as 5BZQ\,J0505+0459, an LBL/FSRQ 
at $z=0.954$), and 2WHSP\,J050833.3+05310 (an HBL), i.e. sources no. 5, 1, 
and 3 respectively in 
Fig.\,\ref{fig:map1} (right), and four additional blazar candidates are present in the area. The 
first two blazars are also bright $\gamma$-ray emitters (Sect. 
\ref{sec:gamma_emission}). 
Visual inspection of the SED of the other sources allowed us to confirm that 
source no. 4 is a good candidate HBL object, while source 7 is likely a 
cluster of galaxies (due to its extended X-ray emission), source 6 is a steep radio 
spectrum object, and source 2 is a nearby ($z=0.03677$) elliptical galaxy 
showing low luminosity X-ray emission ($L\sim 10^{41}$ erg s$^{-1}$ at 1 keV) that 
could be due to a jet or even to non-nuclear sources.

\subsection{$\gamma$-ray emission near IceCube-170922A} \label{sec:gamma_emission}

The $\gamma$-ray emission near IceCube-170922A is dominated at various times either by TXS\,0506+056 or by PKS\,0502+049 but there are also times when the two sources have roughly equal fluxes, 
as shown in Fig. \ref{fig:fermi55800-5gev}.
To investigate which of the sources dominate the $\gamma$-ray emission during the IceCube-170922A event 
and the neutrino flare period we constructed TS maps for these two periods. We observe the following:
\begin{enumerate}
\item Fig.~\ref{fig:fermi57900-1-5gev} shows the TS maps during the period contemporaneous with and before the IceCube-170922A event.
From the maps it appears that TXS\,0506+056 dominates the photon flux of the region at energies $> 1$\,GeV;
\item during the time of the neutrino flare, on the other hand, 
the situation is different, with PKS\,0502+049 being the brightest source at $E> 1$\,GeV and 
TXS\,0506+056 progressively taking over above 2 and 5\,GeV, as shown in Fig.~\ref{fig:fermi57100-1-5gev}.
\end{enumerate}

We tried to unveil any evidence of $\gamma$-ray emission associated with the neighbour HBL 2WHSP\,J050833.3+05310, which is also within the 90 per cent error contour of IceCube-170922A,  
by conducting a series of eleven unbinned likelihood analyses, covering 9 years of observation with {\it Fermi}-LAT. We set 
energy cuts at E $>$1\,GeV and $>$5\,GeV, integrating over 400 days intervals\footnote{Each bin has 100 days superposed 
to the previous/next bin, meaning we cover the following time windows: MJD 54700 to 55100, 55000 to 55400, 55300 to 
55700... and so on.}. We kept all 3FGL sources in the field, setting both normalizations and photon indexes as free parameters. 
At the 2WHSP\,J050833.3+05310 position an additional power-law source was included in the model. The strongest signature 
for $\gamma$-ray emission was found between MJD 55900 to 56300 for the 1\,GeV energy cut, reaching TS $\sim 5$. This result 
was confirmed by the residual map of the region. 

\subsection{$\gamma$-ray light curves}\label{sec:LC}

\begin{figure}
\includegraphics[width=\columnwidth]{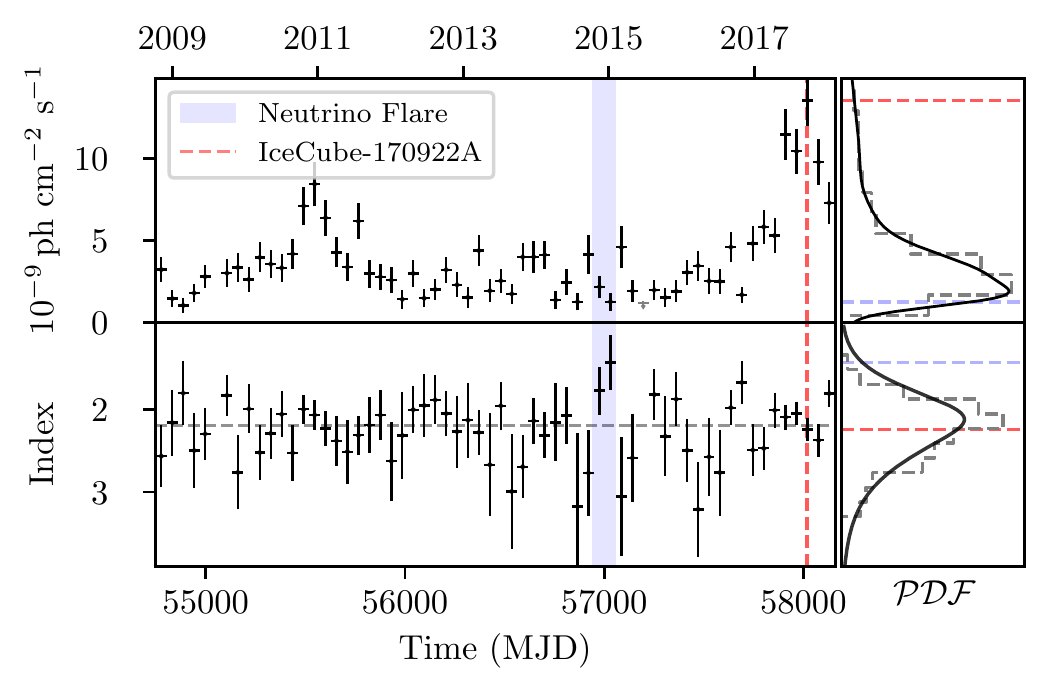}
\includegraphics[width=\columnwidth]{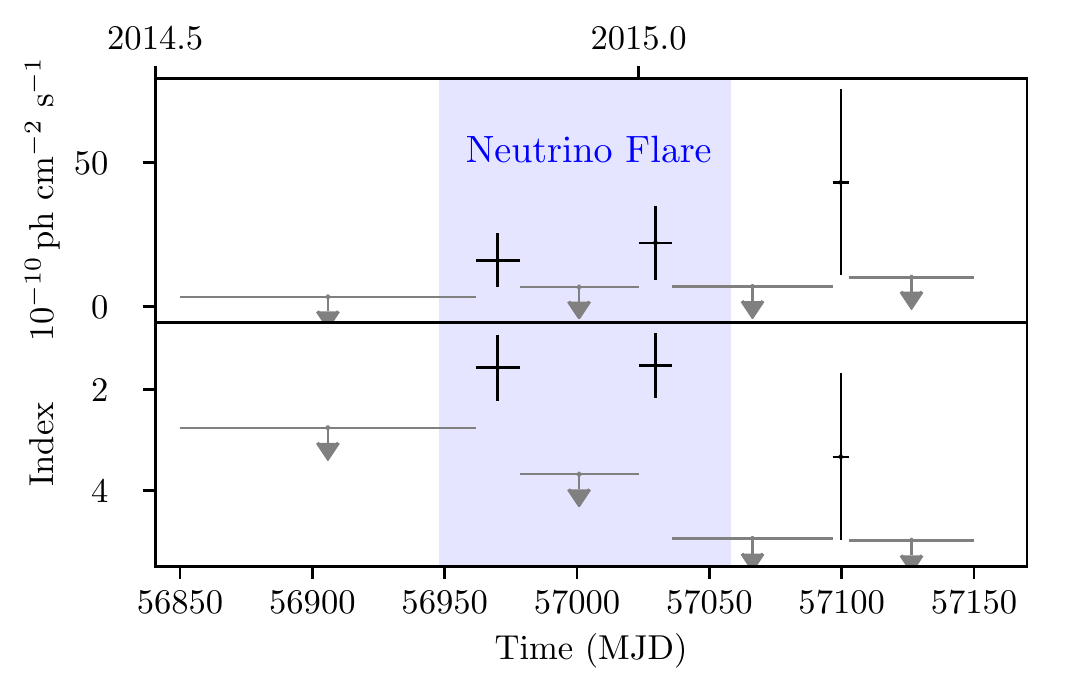}
\caption{{\bf Top}: light curve of TXS\,0506+056 in 55-day bin at $E > 2$\,GeV and photon index curve. All points have 
a TS $> 9$, otherwise 95 per cent upper limits are shown. The blue band denotes the neutrino flare, while the red line 
indicates the IceCube-170922A event. The central horizontal line is the mean 
value of the spectral index. The corresponding photon flux and spectral index mean value
distributions are shown as histograms (dashed grey) and kernel density 
estimations (solid black) in the top- and bottom-right panels respectively, 
with the values during the neutrino flare (blue) and the IceCube-170922A 
event (red) indicated as dashed lines.  
{\bf Bottom}: $\gamma$-ray flux and photon index of TXS\,0506+056 above 10\,GeV, at the time of the neutrino excess observed around MJD\,57000. The photon fluxes are calculated up to 300\,GeV.
}
\label{fig:LC_5BZBJ0509}
\end{figure}

\begin{figure}
\includegraphics[width=\columnwidth]{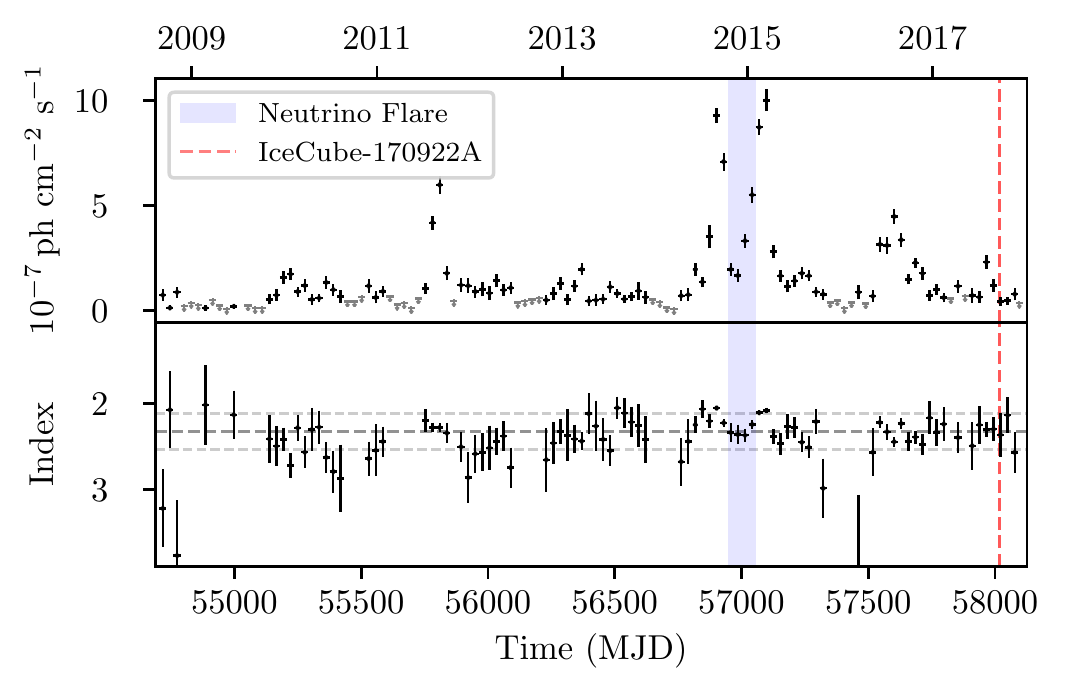}
\includegraphics[width=\columnwidth]{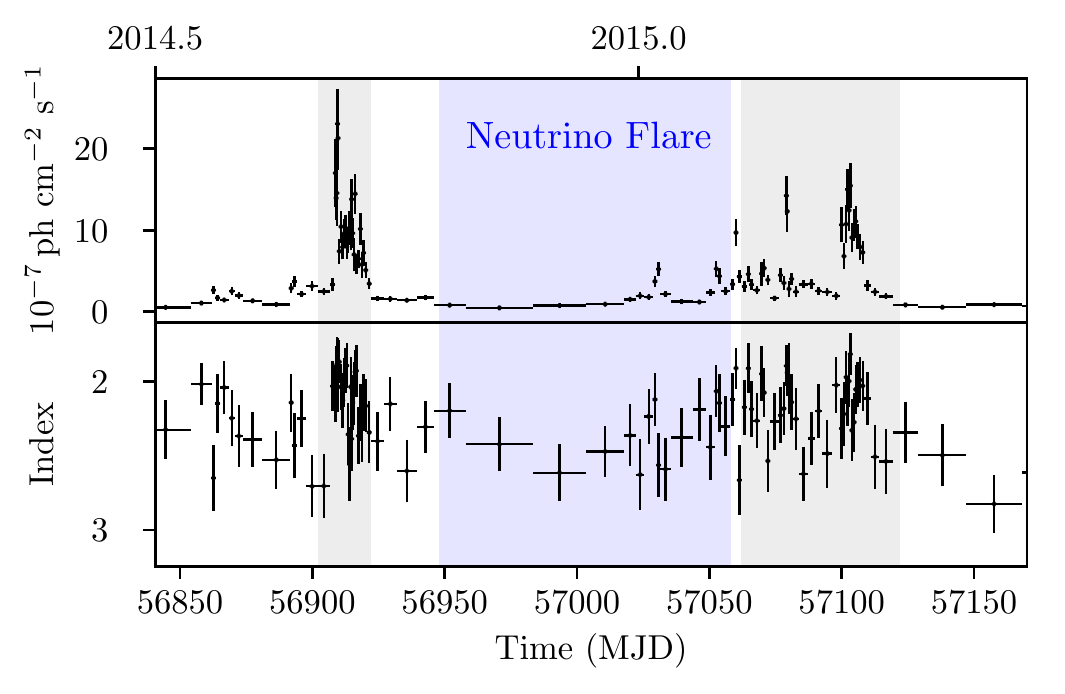}
\caption{{\bf Top}: light curve of PKS\,0502+049 in 28-day bin and $E > 0.1$\,GeV and photon index curve. 
All points have a TS $> 9$, otherwise 95 per cent upper limits are shown. The blue band 
denotes the neutrino flare, while the red line indicates the IceCube-170922A
event. The central horizontal line is the mean value of the spectral index. 
{\bf Bottom}: adaptive bin light curve photon flux with $E_{\rm min} =214$\,MeV $\gamma$-ray light curve of PKS\,0502+049 at the time of the neutrino excess observed around MJD\,57000. 
The photon fluxes are calculated up to 300\,GeV. The shaded bands at the left and right of the neutrino flare indicate the time windows where PKS\,0502+049 is in a high photon flux state (see also Fig. \ref{fig:SED-BZQ}).}
\label{fig:LC_5BZQ}
\end{figure}

As discussed in Sect.\,\ref{sec:lc_pi}, we have derived the $\gamma$-ray light and spectral index 
curves of TXS\,0506+056 and PKS\,0502+049.
As inferred from Fig.\,\ref{fig:fermi57100-1-5gev}, we have evidence that PKS\,0502+049 
dominates the $\gamma$-ray sky at low energies during the neutrino flare, possibly contaminating TXS\,0506+056 below 2\,GeV. 
We build the light curves for TXS\,0506+056 integrating photons above 2\,GeV 
to avoid any bias from PKS\,0502+049 
and also study the highest energies 
(above 10\,GeV) during the specific neutrino flare period.
This is the 
best compromise for the energy threshold. Ideally, one would like to sample as high energies as possible,
to profit from the smaller {\it Fermi}-LAT PSF and source containment region and 
therefore reduce
contamination from PKS\,0502+049. However, the larger the energy, the smaller the 
photon statistics.

In contrast, we build light curves for PKS\,0502+049 above 0.1\,GeV with fixed time 
bins and above 214\,MeV with the adaptive binning method to study both long and short 
term structures. 
Fig.~\ref{fig:LC_5BZBJ0509} (top) shows the light curve of TXS\,0506+056. The 
blue band denotes the neutrino flare, while the red line indicates the 
IceCube-170922A event. Significant flux variations are visible, as typical 
of blazars. The corresponding photon flux and spectral index distributions are 
also shown on the right. It is interesting to note that TXS\,0506+056 was at its hardest in 
the {\it Fermi}-LAT band during the 
neutrino flare, while being relatively faint (a ``low/hard'' state), and at its 
brightest during the 
IceCube-170922A event, while being softer (a ``high/soft'' state). 
Note that, based on the overall distributions, a 
spectral index as hard as observed during the neutrino flare is 
expected with a probability of only $\sim$ 2\ per cent, while a flux as high 
at that during the IceCube-170922A event has a probability of only $\sim$ 1\ 
per cent to be detected. The average photon 
index during the entire duration of the neutrino flare for $E > 
2$\,GeV is $1.62 \pm 0.20$.  

\begin{figure}
\includegraphics[height=9cm]{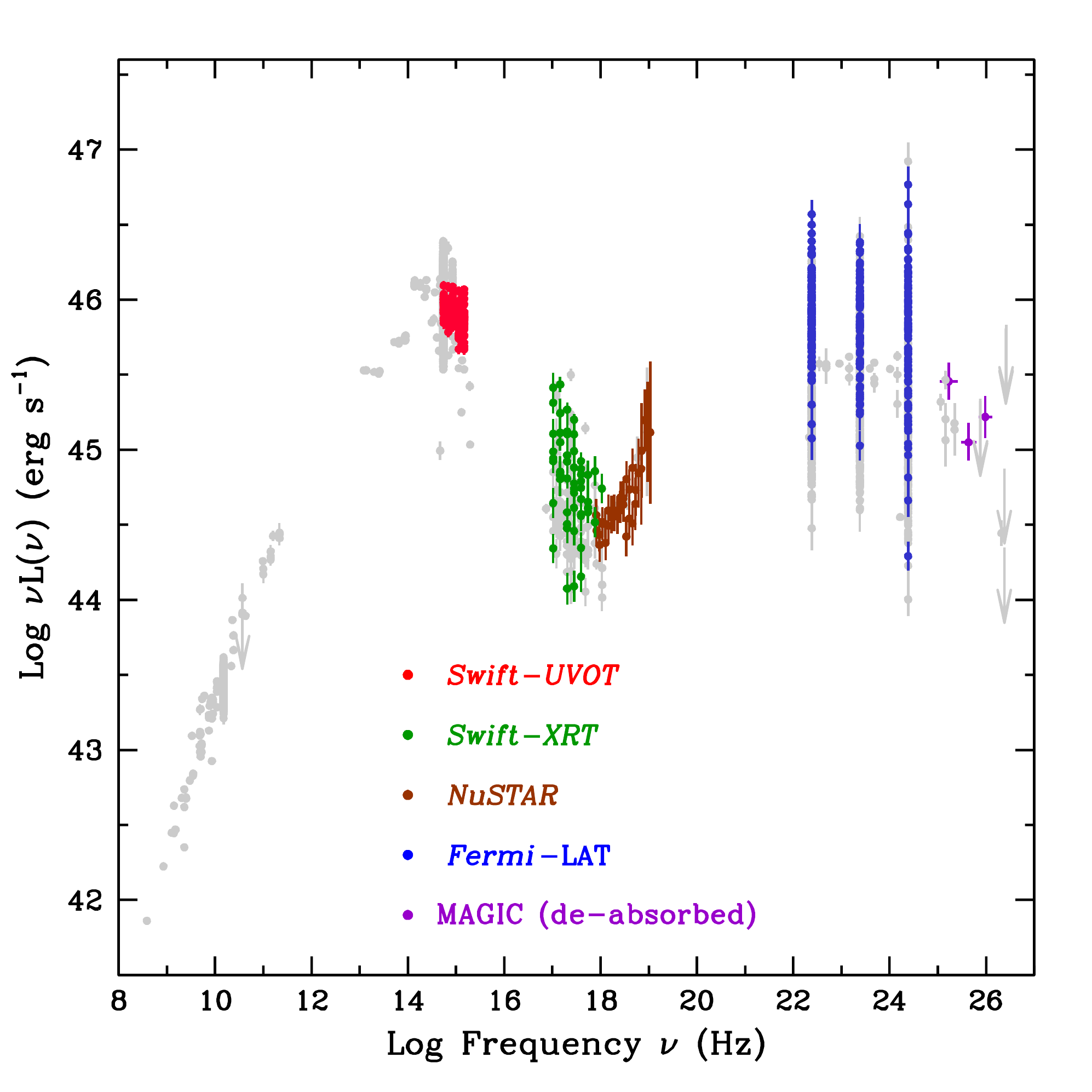}
\caption{The SED of  TXS\,0506+056 assembled using all available public data. The blue measurements in the $\gamma$-ray band show the
variability at 0.1, 1, and 10\,GeV using {\it Fermi}-LAT data collected between 
January 1 and December 31, 2017 (see Fig. \ref{fig:mwlc} for the 1 and 10\,GeV 
light curves). Green and brown points at X-ray frequencies are from our analysis of \swift~and {\it NuSTAR} data respectively. 
MAGIC data \citep{icfermi} are shown as purple crosses and were de-absorbed to correct for the extragalactic background light following 
\protect\cite{Dom_2011}. Red points at optical and UV frequencies are from \swift-UVOT. 
The other non-simultaneous multi-frequency measurements are from catalogues and online archives (grey points).}
\label{fig:SED-BZB}
\end{figure}

\begin{figure}
\includegraphics[height=9.0cm]{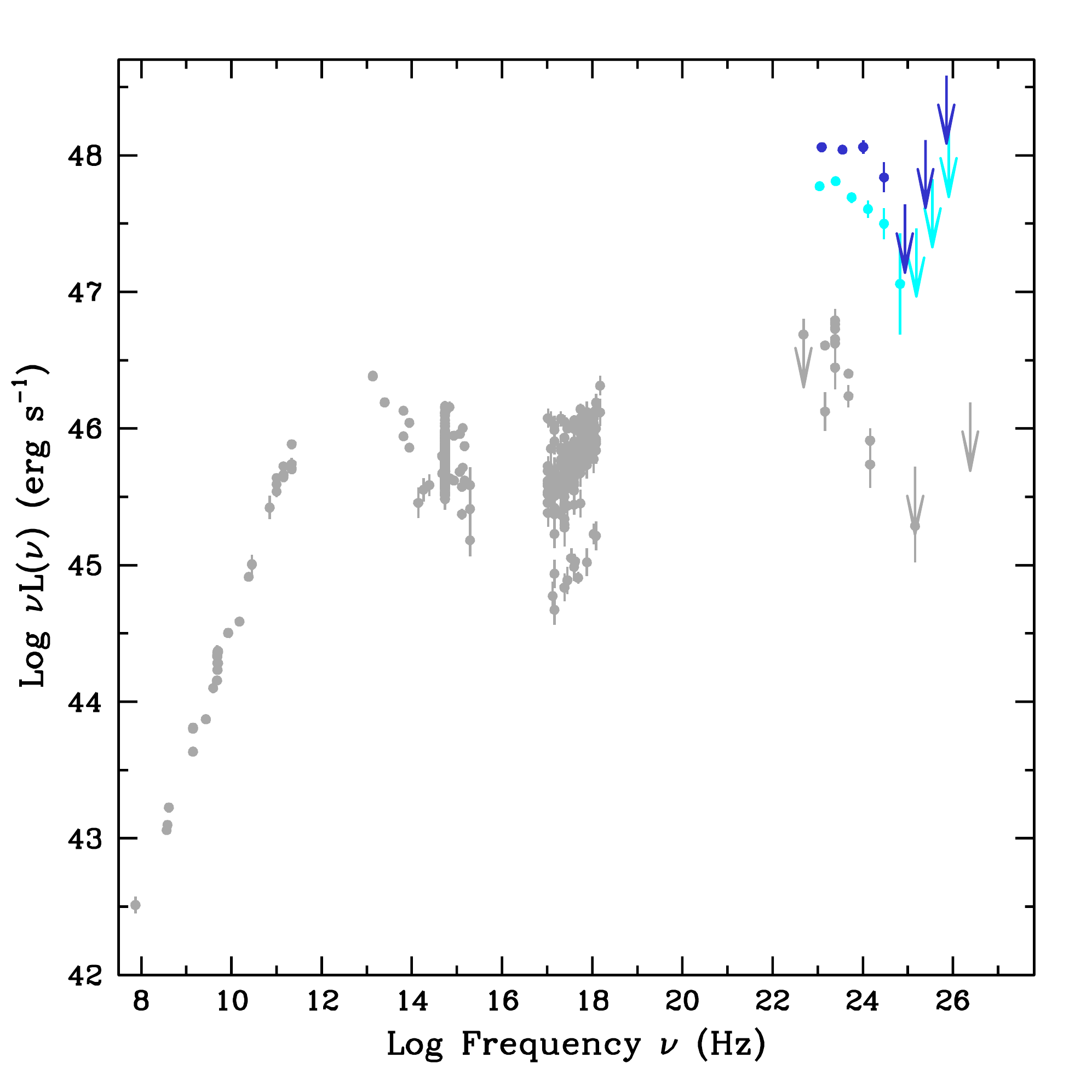}
\caption{The SED of PKS\,0502+049. The $\gamma$-ray spectra during the two flaring periods (see Fig. 
\ref{fig:LC_5BZQ}, bottom) are shown as 
blue (MJD\,56902 -- 56922) and cyan (MJD\,57062 -- 57122) points. All other multi-frequency data are 
non-simultaneous archival data.}
\label{fig:SED-BZQ}
\end{figure}

Since we want to concentrate on the highest energy photons we zoom 
in on the period around the neutrino flare to investigate in 
more detail the source variability looking for the most extreme 
emission using only events above 10 GeV. 
The period from MJD 56850 to MJD 56750 was then divided into half- 
and one-day bins and an unbinned maximum likelihood analysis 
was performed. Next, the nearby bins with TS $> 0$ 
were merged and new light curves were calculated. In order to improve 
the 
statistics, the length of the time periods with TS $> 0$ was then progressively 
increased by adding 1-hour intervals. As a result, we find two periods with significant emission above 10\,GeV 
(Fig.~\ref{fig:LC_5BZBJ0509}, bottom): MJD\,56961.75 -- 56978.29 (TS\,=\,30.5),
with the highest energy photon at 53.3\,GeV, and  
MJD\,57023.25 -- 57036.0 
(TS\,=\,33.6), 
with the highest energy photon at 52.6\,GeV\footnote{Based on the \texttt{gtsrcprob}
tool both of these photons have a $> 99$ per cent probability of being related
to TXS\,0506+056}. 

Fig. \ref{fig:LC_5BZQ} shows the light curve of PKS\,0502+049. The blue band 
denotes the neutrino flare, while 
the red line indicates the IceCube-170922A
event. Significant flux variations are also visible in this case in particular in 
the periods right before and 
after the neutrino flare but the overlap with the neutrino flare is minimal even taking into account
the uncertainties on its duration as clearly 
visible in the adaptive bin zoom-in 
(see Fig. \ref{fig:LC_5BZQ} bottom; the neutrino flare covers 110$^{+35}_{-24}$ days 
and we did not consider small $\gamma$-ray fluctuations
over much shorter time scales). PKS\,0502+049 presents no particular states 
in photon flux and/or spectral index during  IceCube-170922A and the neutrino 
flare period.

\subsection{Spectral energy distributions}

\begin{figure*}
\includegraphics[height=7.3cm]{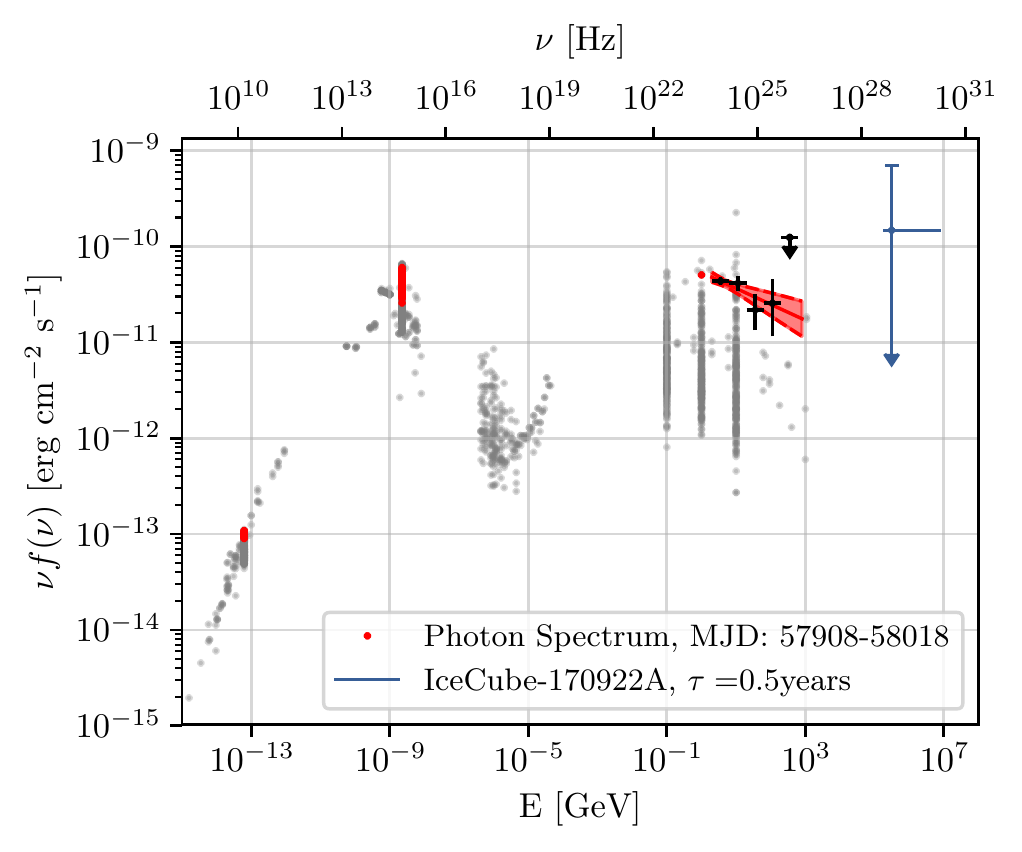}
\includegraphics[height=7.3cm]{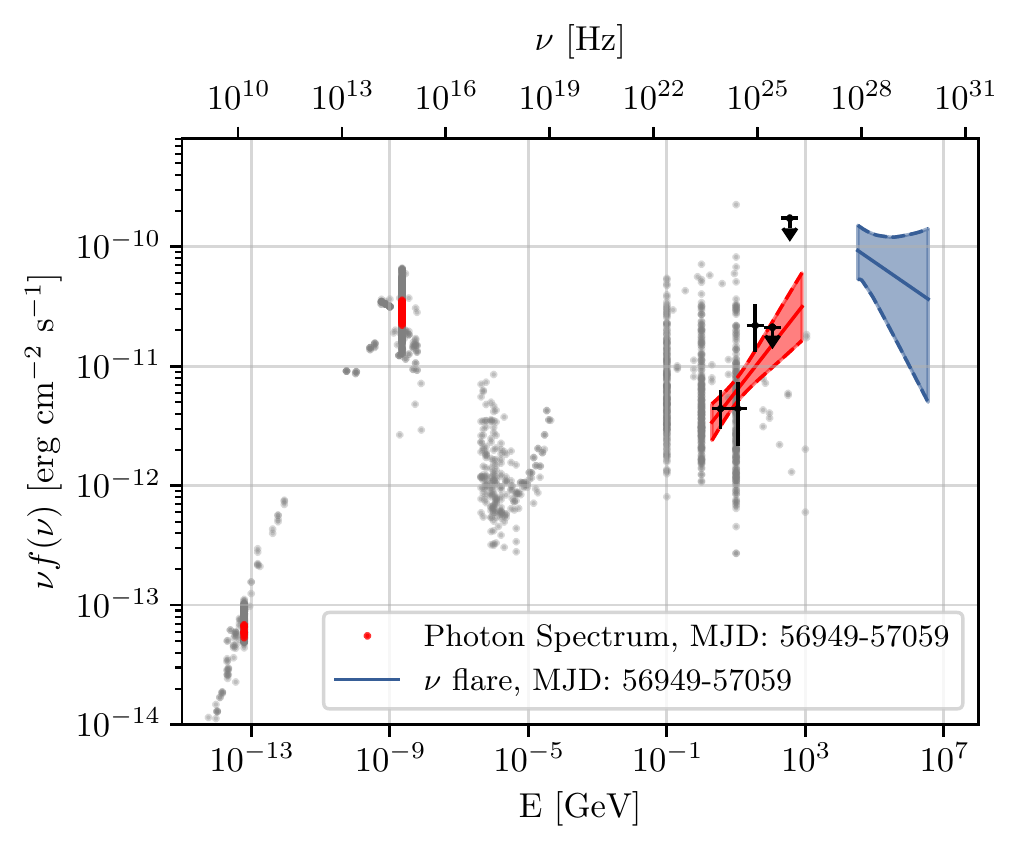}
\caption{The hybrid photon -- neutrino SED of TXS\,0506+056. The red points 
(OVRO at 15 GHz and ASAS V$_{\rm mag}$) are simultaneous 
with neutrinos, grey ones refer to historical data, while the black 
ones are {\it Fermi} data. 
The red bands for the $\gamma$-ray flux 
show the $1\,\sigma$ error bounds on the best fit, while upper 
limits are given at 
95 per cent C.L.. {\it Fermi} data points were de-absorbed to 
correct for the 
extragalactic background light following \protect\cite{Dom_2011}. 
{\bf Left:} the MJD\,57908 -- 58018 period 
(June 4 -- September 22, 2017). The neutrino flux has been derived 
by \protect\cite{icfermi} over the 200\,TeV -- 7.5\,PeV range (see 
text for more 
details); we give here the all-flavour flux. The vertical upper 
limit is drawn at the most probable neutrino energy. The average 
{\it Fermi}-LAT photon index for $E > 
2$\,GeV is $2.16\pm 0.10$. {\bf Right:} the MJD\,56949 -- 57059 
period (October 19, 2014 -- February 6, 2015). The neutrino flux has been 
derived by \protect\cite{iconly} over the 32\,TeV -- 3.6\,PeV range; 
the error is the combined error on the spectral index and the normalization. 
The average {\it Fermi}-LAT photon index for $E > 
2$\,GeV is $1.62 \pm 0.20$.  
}
\label{fig:hybrid_SED}
\end{figure*} 

Fig. \ref{fig:SED-BZB} shows the SED of TXS\,0506+056.  
The {\it NuSTAR} data show a hardening at $\sim 
10^{18}$~Hz ($\sim 4 - 5$~keV) most likely due to the onset of the inverse 
Compton component \citep[e.g.][]{Padovani_2017}. 
Fig. \ref{fig:SED-BZQ} shows the SED of PKS\,0502+049. Two points can be made: (1) during flares
PKS\,0502+049 is a brighter $\gamma$-ray source than TXS\,0506+056; (2) the SED
of PKS\,0502+049 cuts off at $E \approx 10$\,GeV, while that of TXS\,0506+056 reaches
$E \gtrsim 100$\,GeV (compatibly with the likely
extragalactic background light absorption at its redshift: \citealt{Paiano_2018}). 

To study the time evolution of the SED of TXS\,0506+056 we 
have built an animation of nearly simultaneous data, which include 15 GHz monitoring 
data from OVRO, 
the ASAS V$_{\rm mag}$ light curve \citep{asas}, the optical/UV data of the \swift-UVOT \citep{Roming} 
analyzed with the SSDC online tool, and the X-ray 
and $\gamma$-ray data described in sect \ref{sec:X-rayData} and \ref{sec:Gamma-rayData}. 
The animation is available here: \url{https://youtu.be/lFBciGIT0mE}.

\subsubsection{The hybrid SED of TXS\,0506+056}\label{sec:hybrid_TXS}

Fig. \ref{fig:hybrid_SED} shows the hybrid photon -- neutrino SED of TXS\,0506+056 for the period
around the IceCube-170922A event (left) and the neutrino flare (right), based 
on the concept introduced by \cite{Pad_2014}. 
The red points are the electromagnetic emission simultaneous with the neutrinos. 
The detection of high-energy neutrinos above $\sim 30$\,TeV 
implies the existence of protons up to at least $3 \times 10^{14} - 3 \times 10^{15}$\,eV, which then collide with other
protons ($pp$ collisions) or photons ($p\gamma$ collisions). High-energy
 $\gamma$-rays with energy and flux about a factor two higher than the
 neutrinos at the source are then expected
 as secondary products in both cases \citep{2006PhRvD..74c4018K,
 2008PhRvD..78c4013K}.
Indeed, in both cases the (linearly extrapolated) $\gamma$-ray and neutrino fluxes are comparable, consistently
with the hypothesis that they are
produced by the same physical process. 
This is especially true
for the neutrino flare, when the neutrino flux has a relatively small uncertainty 
being derived from $\sim 13$ events within a well-defined period of 110$^{+35}_{-24}$ days 
\citep{iconly}. 

This is different for the IceCube-170922A event, given the large uncertainty 
on the neutrino flux since we are dealing with a single event over an 
ill-defined period of time. To estimate the neutrino flux from one neutrino 
event \cite{icfermi} had to assume a spectral emission shape, an emission 
time $\tau$, and an energy emission range. The corresponding mean number of 
$\nu_\mu$ events $N$ expected in IceCube is 

\begin{align}
N = \tau \int_{E_{\textrm{min}}}^{E_{\textrm{max}}} \textrm{A}_{\textrm{eff}} (E,\theta) \cdot \frac{1}{3}\frac{d\phi}{dE} dE, 
\end{align}

where $\textrm{A}_{\textrm{eff}}$ is the effective area of the IceCube detector and $\frac{1}{3}$ is the flavour ratio assumed.
For a source described by a single power-law distribution 
the flux producing one neutrino event is 

\begin{align}
\phi_0 = \frac{3\cdot N}{\tau \int_{E_{\textrm{min}}}^{E_{\textrm{max}}} \textrm{A}_{\textrm{eff}}
(E,\theta) E^{-\gamma} dE}, 
\end{align}

where $N=1$ and $\tau$ is taken as 0.5 years (of the same order as the duration 
of the $\gamma$-ray flare). Interpreting the observation of one IceCube alert 
event as an upward Poissonian fluctuation, then the flux value calculated 
can be understood as an upper limit on the neutrino flux \citep[see also][]{icfermi}.

\section{Results}

Following up the IceCube-170922A event observed in coincidence with a $\gamma$-ray 
flare of TXS\,0506+056 \citep{icfermi}, the IceCube collaboration has also detected a neutrino flare in late 2014 -- early 2015 from the same direction \citep{iconly}. Given the complexity
of the $\gamma$-ray sky in this area, both spatially and temporally, 
we have carefully dissected the region and found the following: 

\begin{enumerate}

\item out of the 637 radio and X-ray sources within 80 arc-minutes 
of the  
IceCube-170922A event position, only 7 are both radio and X-ray 
emitters and therefore likely
non-thermal sources. As it turns out, the X-ray-to-radio flux ratios of these 7 
sources are blazar-like;
\item out of
these 7 sources, 4 are blazars, two of which are very bright 
$\gamma$-ray sources, namely 
TXS\,0506+056 and PKS\,0502+049, competing for dominance; 
\item while TXS\,0506+056
dominates in all $\gamma$-ray bands during the IceCube-170922A 
event, the situation is more
complex during the neutrino flare, as PKS\,0502+049 dominates up to 
$E \sim 1 - 2$\,GeV but
TXS\,0506+056 takes over at $E \gtrsim 2 - 5$\,GeV. The $\gamma$-ray 
spectrum of PKS\,0502+049,
in fact, cuts off at high energy even during flares, a behaviour 
typical of LBL blazars;
\item PKS\,0502+049 is flaring right before
and right after the neutrino flare (but not in coincidence with it) 
while TXS\,0506+056 was at its
hardest in that time period but in a relatively faint state, 
suggesting a shift to high energies
of the $\gamma$-ray SED; 
\item the hybrid $\gamma$-ray -- neutrino SED of TXS\,0506+056 
during the neutrino flare is as expected for lepto-hadronic models 
since the photon and neutrino fluxes are at the same level
\citep{Petro_2015}. We note that the hybrid 
SEDs of \cite{Pad_2014} and \cite{Padovani_2016} were based on {\it 
one} shower-like IceCube event,
which could in principle have been emitted over the full IceCube 
detection live time, 
and were therefore affected by a very large uncertainty. In the case 
of the neutrino flare, instead, 
a {\it sizable} ($\sim 13$) number of neutrinos has been detected 
within a well-defined time window and good spatial resolution. 

\end{enumerate}

In short, all spatial, timing, and energetic multi-messenger diagnostics point to 
TXS\,0506+056 as the first identified non-stellar neutrino (and therefore cosmic ray) source. 

\section{Discussion}

\subsection{Source properties}\label{sec:properties}

We now explore in more detail the properties of TXS\,0506+056. 
First, we note that this source is a very strong $\gamma$-ray 
source, having an average flux of $7.1 
\times 10^{-8}$ ph cm$^{-2}$ s$^{-1}$ above 100\,MeV, which puts it 
among the top 4 per cent of the {\it Fermi} 3LAC 
catalogue \citep{Fermi3LAC}. Moreover, it also belongs to the 2FHL 
sample \citep{2FHL},
which includes all sources detected above 50\,GeV by {\it Fermi}-LAT 
in 80 months of data.
TXS\,0506+056 also has a large radio flux density $\sim$ 1 Jy at 6 
cm \citep{Gregory_1991}, and 
$\sim 537$ mJy at 20 cm, which makes it one of the brightest radio 
sources (in the top 0.3 per cent) 
of the NRAO VLA Sky Survey, which covers 82 per cent of the sky 
\citep{Condon_1998}. Fig. \ref{fig:SED-BZB} shows 
the overall SED of the source in luminosity, based on the redshift 
of 0.3365 recently reported by \cite{{Paiano_2018}}. 

The peak luminosities of $\sim 2\times 10^{46}$ erg s$^{-1}$ in the 
synchrotron peak, and almost $10^{47}$ erg s$^{-1}$ at 10\,GeV, 
place this object among the  
most powerful BL Lacs  known, particularly in the high-energy/very 
high-energy $\gamma$-ray band. 
For comparison, the corresponding maximum luminosities ever observed 
in MKN 421 (and PKS\,2155$-$304) 
are  $\sim 4\times 10^{45}$ ($\sim 2\times 10^{46}$) and  $\sim 
1.5\times 10^{45}$ ($10^{46}$) erg 
s$^{-1}$, a factor of $\sim 5$ (1) and $\sim 50$ (10) lower than 
TXS\,0506+056 (Giommi et al., in preparation). What seems to be peculiar in this 
source is the very large luminosity at $\sim 10$\,GeV  compared to 
other similar sources. From the overall SED point of view  
TXS\,0506+056 shows a variability range in the $\gamma$-ray band 
(almost a factor 1,000 at 10\,GeV: see Fig. \ref{fig:SED-BZB}}) 
much larger than that observed at the peak of the synchrotron 
emission. Even during the  large $\gamma$-ray flaring event observed 
close to the detection of IceCube-170922A the peak 
of the synchrotron emission (located in the UV band) did not vary by 
more than a factor of 2, nor did the X-ray flux, at the tail of the 
synchrotron peak, change by a large factor.   
This behaviour is consistent with an excess of hard $\gamma$-ray 
radiation possibly associated with hadronic processes. 

We now posses all the elements to calculate reliably 
the luminosity of a high-energy neutrino source. 
Using the 
fluence, spectral index, and energy range 
given in Sect. \ref{sec:nu_flare} and \cite{iconly} 
we 
do the following: 1. derive an integrated $\nu_\mu$ flux of 
$1.2 \times 10^{-10}$ erg cm$^{-2}$ s$^{-1}$ from the fluence
by integrating over the $2\,\sigma$ range around the central
value of the time period; 2. estimate 
$L_{\nu_{\mu}}$; 3. derive a neutrino luminosity all-flavour 
(assuming $\nu_{\rm e}:\nu_\mu:\nu_\tau = 1:1:1$) 
by multiplying by 3 the $\nu_\mu$ power. The result is
$L_{\nu} = 3 \times L_{\nu_{\mu}} \sim  3 \times 4.5 \times 
10^{46}$ erg s$^{-1}$ $\sim 1.4^{+0.6}_{-0.5} \times 10^{47}$ erg 
s$^{-1}$ between 32\,TeV and 3.6\,PeV. (This luminosity is 
fully consistent with the one derived by \cite{iconly} of 
$1.2^{+0.6}_{-0.4} \times 10^{47}$ erg 
s$^{-1}$ based on a flare duration of 158 days derived from the box 
time-window result.)

This can be compared to the {\it simultaneous} $\gamma$-ray 
luminosity $L_{\gamma}$ (2\,GeV -- 1\,TeV)\footnote{This is based on 
the {\it Fermi}-LAT best fit extrapolated to 1\,TeV.} $\sim  3 
\times 10^{46}$ erg s$^{-1}$ [or $L_{\gamma}$ (2\,GeV -- 100\,GeV) 
$\sim 10^{46}$ erg s$^{-1}$]. 

\subsection{Physical implications}

Our results have several physical implications. Namely: 

\begin{figure}
\includegraphics[height=6.3cm]{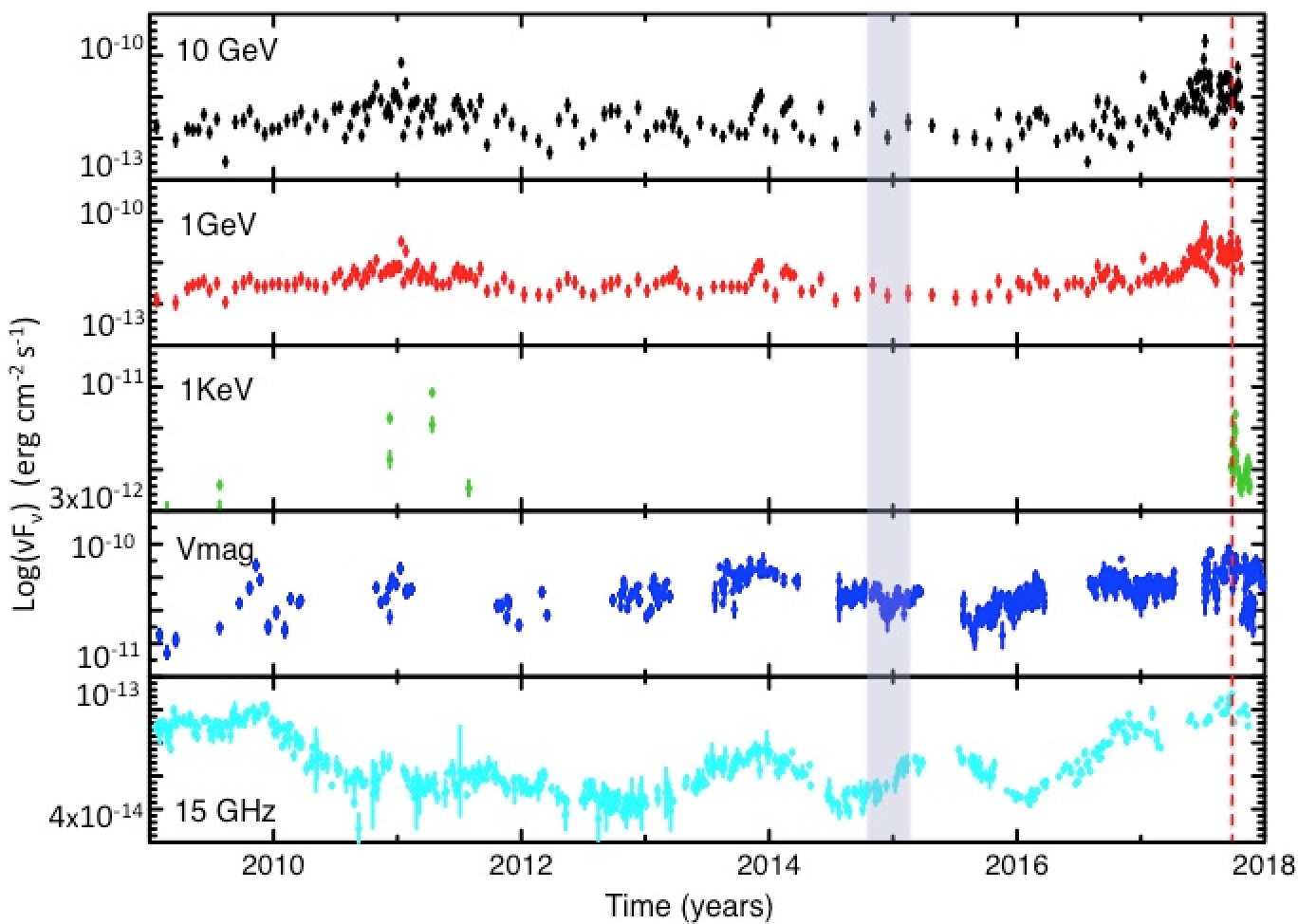}
\caption{The radio (15 GHz), optical (V$_{\rm mag}$), X-ray and $\gamma$-ray light curves of 
TXS\,0506+056. The radio data have been taken from the OVRO database, the visual magnitude 
data are from the Catalina Real Time Transient Survey (CRTS) and from the All Sky Automatic 
Survey ASAS \citep{asas}. The $\gamma$-ray light curves have been produced using {\it Fermi}-LAT 
data with the adaptive-bin method \citep{Lott_2012}. The blue band denotes the neutrino flare, while 
the red line indicates the IceCube-170922A event.}
\label{fig:mwlc}
\end{figure}

\begin{enumerate}
\item with neutrinos we are exploring an energy range which is inaccessible with photons 
at this redshift. Therefore, the first evidence for neutrino emission from the direction of 
TXS\,0506+056 opens a new window on blazar physics;
\item the derived $L_{\nu}$ during the neutrino flare is quite large and even larger than  
$L_{\gamma}$ (2\,GeV -- 1\,TeV), which might imply that a sizeable fraction of the 
neutrino-related $\gamma$-rays have energies above the {\it Fermi}-LAT energy band,
as expected in the case of $p\gamma$ collisions
(Sect. \ref{sec:hybrid_TXS});
\item  the ratio between the two SED humps (high-energy to low-energy, usually known as Compton dominance [CD]: e.g. 
\citealt{planck}) for 
TXS\,0506+056 is $>1$, while typical HBL have CD $\sim$ 0.1. This could be due to the 
electromagnetic emission coming from different blobs e.g. one dominant at optical 
frequencies and the other dominant at high-energy $\gamma$-rays, or to the presence of an
additional hadronic component in the $\gamma$-ray band;
\item a discrete cross-correlation analysis using the z-transformed discrete correlation function \citep[ZDCF;][]{zdcf}
among  the radio (15 GHz), optical, X-ray and $\gamma$-ray light curves of TXS\,0506+056, was built with the data 
described in the previous paragraphs (Fig. \ref{fig:mwlc}). The ZDCF  
shows a strong correlation ($\sim 10\,\sigma$) with a time lag of $\approx 4$ months between the radio and 
optical emission (with the optical band leading) but no correlation between optical/X-ray/radio and $\gamma$-ray bands.
This is at variance with typical single-zone leptonic (synchrotron self-Compton 
or external 
Compton) models where all energy bands are expected to be well correlated;
\item the SED of at least one blazar now {\it has} to be modeled within a lepto-hadronic 
scenario \citep[e.g.][]{Petro_2015}. This is far from trivial and goes well beyond the scope of this
paper;
\item the neutrino sky has been populated so far by only two sources: the Sun \cite[e.g.][]{2014Natur.512..383B} and 
SN 1987A \citep{1987PhRvL..58.1490H,1987PhRvL..58.1494B,1987PZETF..45..461A}. TXS\,0506+056 
is now a third plausible candidate, whose neutrino energies are, however, more than six orders of magnitude
larger than those of the two stellar sources. 
\end{enumerate}

\subsection{Future searches}
Our results suggest two periods of neutrino emission for TXS\,0506+056, which appear to 
be in connection with two very different 
$\gamma$-ray states, namely one high/soft, connected to the IceCube-170922A event, and 
another one low/hard, related to the neutrino flare. This implies that the search 
for multi-messenger sources needs to be carried out not only for flaring $\gamma$-ray sources 
but also for relatively hard emitters. These criteria can also be used by neutrino telescopes to look for other neutrino sources.

Why is this the only strong IceCube neutrino source candidate? What 
makes it special? We believe
this is due to a series of factors. First, as discussed in Sect. 
\ref{sec:properties},
TXS\,0506+056 is a very strong $\gamma$-ray and radio source. 
Moreover, its declination
and energy range happened to be in the regions of parameter space 
where IceCube reaches  
maximum sensitivity and the duration of the flare was also long 
enough for IceCube to 
(marginally) detect it (see also \citealt{iconly}).  

Future searches for cosmic neutrino sources should emphasize: (1) extreme blazars
of the BL Lac type, as hinted at by \cite{Padovani_2016}; (2) high/soft -- low/hard $\gamma$-ray 
states; (3) regions of parameter space where the neutrino detectors are most sensitive. 

\section{Conclusions}

The IceCube-170922A event and the neutrino flare at the end of 2014 have been 
linked
to the same source, TXS\,0506+056, a blazar of the BL Lac type at $z = 0.3365$. This is the 
most plausible 
association so far between IceCube neutrinos and an extragalactic object.  
The area near TXS\,0506+056 is quite complex due to the presence of several non-
thermal
sources, which in principle could all contribute to the overall $\gamma$-ray flux. 
We have therefore carefully dissected this region using a multi-messenger 
approach, 
obtaining the following results:

\begin{enumerate}
\item TXS\,0506+056 was the brightest {\it Fermi} 
source in the region of interest at energies above 1\,GeV 
during the IceCube-170922A event but only above $2 - 5$\,GeV during
the neutrino flare. PKS\,0502+049, a nearby blazar of the FSRQ type offset by $\sim 1.2^{\circ}$, 
dominated the $\gamma$-ray region at lower energies in the latter period. 
\item We have observed two periods of significant neutrino emission consistent 
with the position of TXS\,0506+056
in connection with two very different $\gamma$-ray states, namely one, high/soft, 
connected to the IceCube-170922A event, 
and another one, low/hard, related to the neutrino flare. PKS\,0502+049 was also 
flaring right before
and right after the neutrino flare but not in coincidence with it. 
\item We have built the hybrid photon -- neutrino SED of TXS\,0506+056 during 
the neutrino flare and have reliably estimated the power of a high-energy 
neutrino source.
This is $\sim 1.4 \times 10^{47}$ erg s$^{-1}$ 
between 32\,TeV and 3.6\,PeV (all-flavour), 
even larger than the {\it simultaneous} $L_{\gamma}$ (2\,GeV -- 1\,TeV) $\sim 3 
\times 10^{46}$ erg s$^{-1}$.
\item Both the lack of a correlation between the $\gamma$-ray and radio/optical 
flux and the 
SED shape of TXS\,0506+056, which is unusual in terms of its Compton dominance, appear not to be 
consistent with 
simple leptonic models. 
\item All of the above is fully consistent with the hypothesis that TXS\,0506+056 
has undergone a 
hadronic flare during the neutrino detections.
\end{enumerate}

In short, all spatial, timing, and energetic multi-messenger diagnostics point to 
TXS\,0506+056 as the only counterpart of all the neutrinos observed in the 
vicinity of IceCube-170922A and therefore the first non-stellar neutrino (and hence cosmic ray) 
source. The emergent picture is that extreme blazars, i.e., strong, very high energy 
$\gamma$-ray sources with the peak of the synchrotron emission $> 10^{14} - 10^{15}$
Hz
\citep{Padovani_2016}, are the first class of 
sources with evident contribution to the IceCube diffuse signal 
\citep{ICECube17_1,ICECube17_2}. 

Future searches for cosmic neutrino sources, concentrated on similar classes of sources, using 
additional high-energy track-like events, and based on detailed multi-messenger analysis, 
will likely provide further associations. 

\section*{Acknowledgments}

We acknowledge the use of data and software facilities 
from the SSDC, managed by the Italian Space Agency,
and preliminary versions of other software tools currently 
being developed within the United Nations ``Open Universe'' 
initiative.  PG
acknowledges the support of the Technische Universit{\"a}t M{\"u}nchen -
Institute for Advanced Studies, funded by the German 
Excellence Initiative (and the European Union Seventh Framework 
Programme under grant agreement n. 291763). BA is supported by the 
S\~ao Paulo Research Foundation (FAPESP) with grant n. 2017/00517-4. 
This work is supported by the Deutsche Forschungsgemeinschaft through grant SFB\,1258 
``Neutrinos and Dark Matter in Astro- and Particle Physics''.
We thank the IceCube collaboration for granting us access to information
prior to publication, through a Memorandum of Understanding. We thank Markus Ahlers, Dave Seckel, and Justin Vandenbroucke for fruitful discussions
and the referee, Dr. Ralph Wijers, and the scientific editor for useful and swift comments.
The \textit{Fermi}-LAT Collaboration acknowledges generous ongoing support
from a number of agencies and institutes that have supported both the
development and the operation of the LAT as well as scientific data analysis.
These include the National Aeronautics and Space Administration and the
Department of Energy in the United States, the Commissariat \`a l'Energie Atomique
and the Centre National de la Recherche Scientifique / Institut National de Physique
Nucl\'eaire et de Physique des Particules in France, the Agenzia Spaziale Italiana
and the Istituto Nazionale di Fisica Nucleare in Italy, the Ministry of Education,
Culture, Sports, Science and Technology (MEXT), High Energy Accelerator Research
Organization (KEK) and Japan Aerospace Exploration Agency (JAXA) in Japan, and
the K.~A.~Wallenberg Foundation, the Swedish Research Council and the
Swedish National Space Board in Sweden. Additional support for science analysis during the 
operations phase is gratefully acknowledged from the Istituto Nazionale di Astrofisica in Italy and 
the Centre National d'\'Etudes Spatiales in France.


\label{lastpage}

\bsp	

\end{document}